\documentclass[%
 reprint,
superscriptaddress,
 amsmath,amssymb,
 aps,
 prl,
]{revtex4-1}

\usepackage{graphicx}
\usepackage{dcolumn}
\usepackage{bm}

\begin{document}


\title{Dressed Rabi oscillation in a crystalline organic radical}
\author{C. Blake Wilson}
\affiliation{Department of Physics, University of California, Santa Barbara, Santa Barbara, California, USA}
\affiliation{Institute for Terahertz Science and Technology, University of California, Santa Barbara, Santa Barbara, California, USA.}

\author{Devin T. Edwards}
\affiliation{Department of Physics, University of California, Santa Barbara, Santa Barbara, California, USA}
\affiliation{Institute for Terahertz Science and Technology, University of California, Santa Barbara, Santa Barbara, California, USA.}

\author{Jessica A. Clayton}
\affiliation{Department of Physics, University of California, Santa Barbara, Santa Barbara, California, USA}
\affiliation{Institute for Terahertz Science and Technology, University of California, Santa Barbara, Santa Barbara, California, USA.}

\author{Songi Han}
\affiliation{Institute for Terahertz Science and Technology, University of California, Santa Barbara, Santa Barbara, California, USA.}
\affiliation{Department of Chemistry and Biochemistry, University of California, Santa Barbara, Santa Barbara, California, USA}

\author{Mark S. Sherwin}
\affiliation{Department of Physics, University of California, Santa Barbara, Santa Barbara, California, USA}
\affiliation{Institute for Terahertz Science and Technology, University of California, Santa Barbara, Santa Barbara, California, USA.}

\date{\today}

\begin{abstract}
Free electron laser-powered pulsed electron paramagnetic resonance (EPR) experiments performed at 240 GHz / 8.56 T on the crystalline organic radical BDPA reveal a tip-angle dependent resonant frequency. Frequency shifts as large as 11 MHz (45 ppm) are observed during a single Rabi oscillation. We attribute the frequency shifts to a ``dressing" of the nutation by spin-spin interactions. A nonlinear semi-classical model which includes a temperature- and sample-geometry-dependent demagnetizing field reproduces experimental results. Because experiments are performed without a cavity, radiation damping—the most common nonlinear interaction in magnetic resonance—is negligible in our experiments.

\end{abstract}


\maketitle

Understanding the magnetic properties of systems with unpaired electron spins is at the heart of much of modern condensed matter physics. Over the past decades, extensive efforts have focused on understanding systems at thermal equilibrium. As a result, ordered ground states such as the ferromagnet (FM) and antiferromagnet (AFM) are now well understood. Much current research on magnetism in thermal equilibrium focuses on the search for more exotic, highly entangled ground states such as quantum spin liquids in frustrated magnets \cite{Anderson_Resonating_1973,Senthil_Z2_2000, Moessner_Resonating_2001, Kitaev_Anyons_2006, balents2010spin,Powell_Quantum_2011, Savary_2016, Vojta_2018}. The behavior of large numbers of interacting spins far from equilibrium, however, remains challenging to study despite intense theoretical \cite{Floquet_time_crystals} and experimental investigations \cite{choi2017observation}.

Crystalline organic radicals\textemdash crystals composed of organic molecules that each contain an unpaired electron spin\textemdash are promising candidates for hosting interesting phases of strongly interacting spins driven far from equilibrium. Spins are sufficiently close to each other that the dominant interaction is exchange, but, because of small spin-orbit coupling, spins are sufficiently isolated from the lattice that a non-equilibrium spin state decays much more slowly than the time scales associated with spin-spin interactions.

BDPA (1,3-bisdiphenylene-2-phenylallyl), also known as the Koelsch radical, is an organic spin-1/2 radical that has been the subject of much study \cite{Duffy_Antiferromagnetic_1972, Azuma_Molecular_1994, Mitchell_Electron_2011, Yamauchi_Low_1977, Hamilton_Linear_1963} and is widely used as a standard sample in electron paramagnetic resonance (EPR) \cite{Bennati_Pulsed_1999, Goldfarb_HYSCORE_2008, Durkan_Electronic_2002} and as a polarizing agent in dynamic nuclear polarization (DNP) enhanced nuclear magnetic resonance (NMR) experiments \cite{Dane_Synthesis_2010, Giraudeau_Multiple_2009}. Crystallized 1:1 complexes of BDPA with benzene (BDPA-Bz, C$_{39}$H$_{27}$) are well described by a quasi-one dimensional Heisenberg AFM linear chain model, with an exchange integral $J/k_B = -4.4$ K along the chain \cite{Duffy_Antiferromagnetic_1972,Yamauchi_Low_1977}, isosceles type spin frustration \cite{Azuma_Molecular_1994}, and the onset of antiferromagnetic order at 1.695 K \cite{Duffy_Antiferromagnetic_1972}. Well above the ordering temperature, the exchange interaction remains important. Thermal fluctuations in the exchange fields narrow the EPR line to a value much smaller than one would expect based on dipole-dipole interactions \cite{Anderson_Exchange_1953, Sushil_Multifrequency_2011}.

 \begin{figure}
    \centering
    \includegraphics{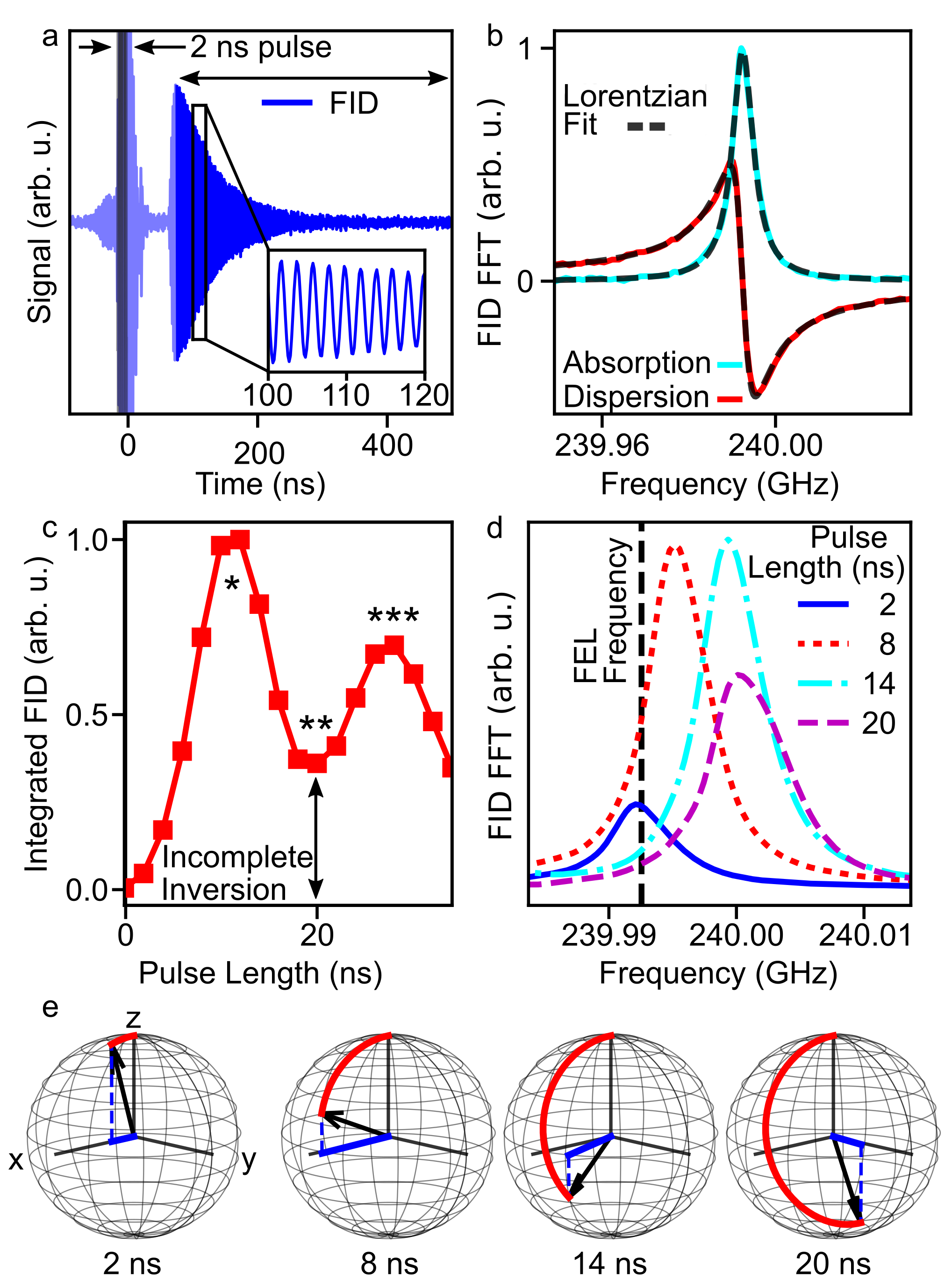}
    \caption{
    \textbf{a} Room-temperature FID signal generated by a 2 ns FEL pulse, highlighted in grey, applied to a BDPA-Bz grain. Inset: the signal digitized at the intermediate frequency (IF = 500 MHz, shown in inset). \textbf{b} Fourier transform of the FID generated by a 2 ns pulse. The FT-EPR lineshape is well described by a Lorentzian with a FWHM of $5.9 \pm 0.1$ MHz. \textbf{c} Integrated FT-EPR intensity as a function of pulse length, demonstrating Rabi oscillations. The magnetic field was chosen so the FID generated by a 2 ns pulse is on resonance with the FEL pulse. For pulse lengths corresponding to the two maxima, the sample magnetization is rotated by $\pi/2$ (*) and by $3\pi/2$ (***), while at the minimum (**) the magnetization has been rotated by $\pi$. The minimum of the Rabi oscillation is not at zero, indicating incomplete population inversion. \textbf{d} FT-EPR absorption lineshape plotted for four pulse lengths. The FID generated by a 2 ns pulse is on resonance with the FEL pulse, FIDs generated by longer pulses are not. \textbf{e} In red, the trajectory taken by the sample magnetization on the Bloch sphere. In blue, the magnetization vector projected onto the x-y plane.
    }
    \label{fig:FT_and_TD_Figure}
\end{figure}

In this Letter, we report the observation of a novel dynamical phenomenon we call dressed Rabi oscillations. Rabi oscillations occur when an ensemble of spins is placed in a static magnetic field $B_0$ and driven by a transverse magnetic field $B_1$ oscillating near the Larmor frequency $\omega_0 = g\mu_B B_0/\hbar$. If the Rabi frequency $\omega_1 = g\mu_B B_1/\hbar$ is larger than the linewidth $\Gamma$ of the magnetic resonance, the spin ensemble undergoes nutation and the direction of the spin ensemble magnetization vector traces out circles on the surface of the Bloch sphere in the rotating frame. We report on Rabi oscillation experiments performed on grains of crystalline BDPA with $\Gamma/2\pi = 4 - 6$ MHz at $B_0 = 8.56$ T, where the electron spin Larmor frequency is 240 GHz, driven by pulses of resonant radiation from the UCSB mm-wave free electron laser with Rabi frequencies $\omega_1$ as large as 25 MHz. We find that Rabi oscillations are dressed by spin-spin interactions and exhibit nonlinear dynamics which we model semi-classically with a Bloch equation modified to include the effects of sample magnetization.

Individual BDPA-Bz grains were mounted on a silver-coated mirror and placed at the end of a corrugated waveguide which tapers to a diameter of 5 mm, then loaded into the center of a tunable (0 to 12.5 T) superconducting magnet. Each grain was $\sim$300 to $\sim$500 $\mu$m across, significantly smaller than the 1.25 mm wavelength 240 GHz radiation. Linearly polarized 240 GHz radiation generated by the UCSB mm-FEL in ``long" pulses of 1 to 3 $\mu$s were ``sliced" into excitation pulses of well-defined duration by light-activated silicon switches \cite{hegmann_generation_1996}. The silvered mirror below the BDPA-Bz reflected both the linearly polarized excitation pulse and the circularly polarized EPR signal to a superheterodyne receiver. A light-activated silicon isolation switch placed before the receiver was turned on only after the excitation pulse had passed. For short excitation pulses $<40$ ns, the FEL cavity dump coupler \cite{Takahashi_Cavity_2009} was used to boost the excitation pulse power. When longer excitation pulses were required, the cavity dump coupler was not activated. Single-frequency operation of the FEL was achieved through injection-locking \cite{Takahashi2007}. For details of the FEL-EPR spectrometer, see \cite{Takahashi_Pulsed_2012,Edwards2013, Wilson_MultiStep_2018}.

Figure \ref{fig:FT_and_TD_Figure}a shows the response of a single BDPA-Bz grain placed in a 8.56 T field to a 2 ns long, resonant excitation pulse with a Rabi frequency $\omega_1/2\pi = 25$ MHz. The short, resonant pulse tips the sample magnetization away from thermal equilibrium, which subsequently precesses at 240 GHz and emits circularly polarized magnetic dipole radiation proportional to the magnetization's projection onto the transverse plane. After a $\sim$75 ns delay, the silicon isolation switch activates and a free induction decay (FID) is acquired by the receiver, mixed down to an intermediate frequency (IF = 500 MHz), and digitized. The complex FID signal is then Fourier transformed to extract the Fourier transform-EPR (FT-EPR) lineshape (Figure \ref{fig:FT_and_TD_Figure}b). Figure \ref{fig:FT_and_TD_Figure}c shows the integrated FT-EPR power spectrum as a function of pulse length. As the pulse length increases, the magnetization rotates further in the Bloch sphere, undergoing Rabi oscillations \cite{Rabi_Space_1937}. At the first signal maximum, occurring for a 10 ns excitation pulse duration, the magnetization has rotated by $\pi/2$, while at the first minimum, occurring at around 20 ns, the magnetization has nominally been inverted. However, the fact that this minimum does not correspond to an integrated FT-EPR intensity near zero indicates inversion is far from complete.

Analysis of the FT-EPR signal in frequency space reveals the frequency of the FID, as measured by the FT-EPR signal, changes as a function of pulse length (Figure \ref{fig:FT_and_TD_Figure}d). This indicates that the EPR frequency $\omega_0$, typically given by the Larmor condition $\omega_L = \gamma B$ where $\gamma = g\mu_B/\hbar$, changes as a function of pulse length, a situation unexpected in conventional EPR experiments. The driven change in EPR frequency $\omega_0$ can be attributed to the mean field exerted on one spin by all the other spins in the sample. The dressed Rabi oscillation phase can be modeled in a mean-field sense by modifying the semiclassical Bloch equations to include the demagnetizing field $H_M$ of the BDPA-Bz grain itself, by analogy to ferromagnetic resonance (FMR) \cite{Kittel_Ferro_1948}. For an ellipsoid, $\mathbf{H}_M = -N\cdot\mathbf{M}$ where $\mathbf{M}$ is the magnetization and $N$ is the demagnetization tensor \cite{Zangwill_Electrodynamics_2013}. Writing the total magnetic field $\mathbf{B} = \mu_0(\mathbf{H}_0 + \mathbf{H}_M + \mathbf{M})$, the Bloch equations in the absence of an excitation pulse and neglecting relaxation become
\begin{align}
    \frac{d}{dt}\mathbf{M} = \gamma \mathbf{M} \times \mu_0\big(\mathbf{H}_0 - N\cdot\mathbf{M}\big)
    \label{eq:ModifiedBlochEquations}
\end{align}
where $\mathbf{H}_0 = \mathbf{B}_0/\mu_0$ is the externally applied magnetic field, $\gamma = g\mu_B/\hbar$ is the electron gyromagnetic ratio, and $\mathbf{M}\times\mathbf{M}$ identically vanishes.

Modeling the BDPA-Bz grain as an ellipsoid, the freely precessing magnetization $\mathbf{M}$ obeys the equations of motion
\begin{subequations}
\begin{align}
    \frac{d}{dt}M_x &= \mu_0\gamma\Big(H_0 - (\delta_z - \delta_x)M_z \Big)M_y - M_x/T_2 \\
    \frac{d}{dt}M_y &= -\mu_0\gamma\Big(H_0 - (\delta_z - \delta_x)M_z \Big)M_x - M_y/T_2 \\
    \frac{d}{dt}M_z &= \mu_0\gamma(\delta_x - \delta_y)M_x M_y - (M_z - M_0)/T_1
\end{align}
\label{eq:BlochSpelledOut}
\end{subequations}
where $\delta_x,\delta_x,\delta_z$ are the principal values of $N$, $M_0$ is the equilibrium magnetization, and $T_1$ and $T_2$ are the phenomenological Bloch spin-spin and spin-lattice relaxation times \cite{Bloch_Nuclear_1946}. Assuming axial symmetry so that $\delta_x = \delta_y = \delta_{\perp}$ and $\delta_z = \delta_{\parallel}$, the magnetization precesses at a frequency $\omega(M_z)$ which is a function of the $z$-component of $\mathbf{M}$,
\begin{align}
    \omega(M_z) = \mu_0\gamma(H_0 - \theta_d M_z)
    \label{eq:FrequencyVsMz}
\end{align}
where $\theta_d = \delta_{\parallel} - \delta_{\perp}$. Two special cases are of particular interest: for a very thin, flat disk, $\theta_d = 1$, while for a sphere, $\theta_d = 0$ and the nonlinear term vanishes.

Equations \ref{eq:BlochSpelledOut} and \ref{eq:FrequencyVsMz} predict that the small tip-angle EPR frequency is shifted from the Larmor condition by an amount $\Delta \omega = -\mu_0\gamma\theta_d M_z$, neglecting higher order terms of order $|M_z^2/H_0|$. The dependence of the small tip-angle frequency on the magnetization, through the demagnetizing field, is analogous to the FMR condition which appears in the Kittel equations \cite{Kittel_Ferro_1948}. However, unlike in the FMR case where changes in $M_z$ due to resonant driving fields are negligible, for a paramagnet the full magnetization $\mathbf{M}$ can be rotated into the transverse plane or even inverted by a resonant microwave pulse. From Equations \ref{eq:BlochSpelledOut} and \ref{eq:FrequencyVsMz}, a near-resonance microwave excitation pulse shifts the precession frequency by an amount proportional to the change in $M_z$, with the maximum precession frequency shift $\Delta F = 2\gamma\mu_0\theta_d M_0/2\pi$ occurring after the initial magnetization $M_0$ is inverted. For BDPA-Bz, the maximum frequency shift can be estimated at room temperature in terms of the equilibrium magnetization $M_0 = M_{eq} = 270 A/m$ \cite{Duffy_Antiferromagnetic_1972} to be $\Delta F \sim\theta_d\times 19$ MHz, which is on the order of the observed frequency shifts and several times the observed linewidth (Figure \ref{fig:FT_and_TD_Figure}).

\begin{figure}
    \centering
    \includegraphics{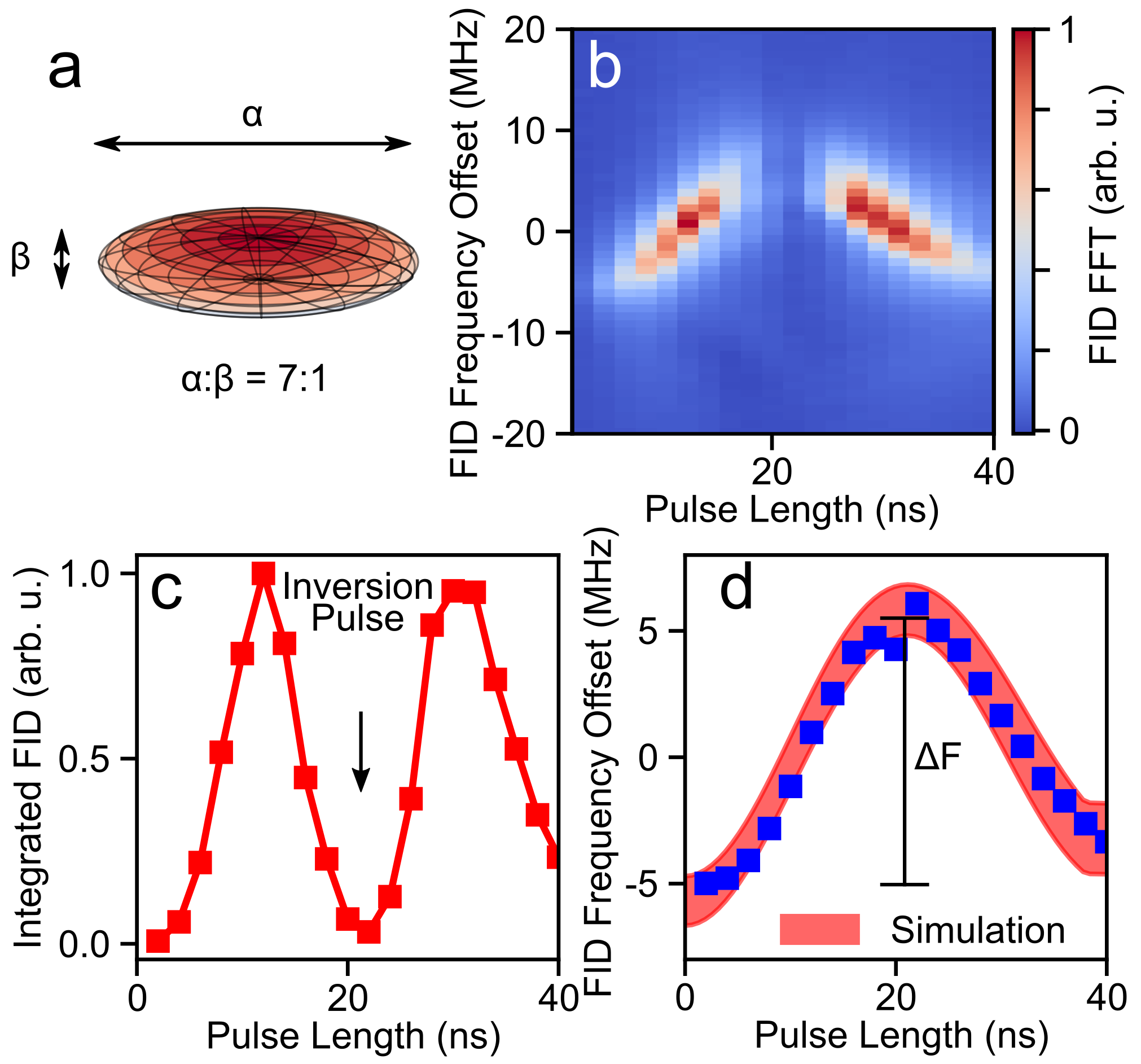}
    \caption{
    \textbf{a} Schematic representation of a BDPA-Bz crystal as an oblate spheroid, with a ratio of major to semi-minor axes of 7:1. \textbf{b} Contour plot of the FT-EPR absorption lineshape, \textbf{c} integrated FID intensity, and \textbf{d} mean FID frequency, as a function of pulse length as BDPA-Bz magnetization undergoes Rabi oscillations. The FID frequency axes are referenced to the FEL frequency. The magnetic field is chosen so that the FID generated by a $\pi/2$ pulse has the same frequency as the FEL. The Rabi oscillations minimum is close to zero, indicating nearly complete population inversion is achieved with the magnetic field thus chosen.
    }
    \label{fig:RabiCurveFigure}
\end{figure}

A grain of BDPA-Bz with a geometry close to an axially-symmetric ellipsoid (Figure \ref{fig:RabiCurveFigure}a), which had an aspect ratio of roughly 7:1 for $\theta_d = 0.65$, was selected for further experiments. Figure \ref{fig:RabiCurveFigure}b shows a nutation experiment performed on this BDPA-Bz grain, with the colorbar indicating the FT-EPR signal strength and with the FT-EPR frequency indicated on the vertical axis. The magnetic field was chosen so the excitation pulses are resonant with the bare Larmor frequency, rather than with the small tip-angle frequency. As a consequence, the spin system initially has a precession frequency below the excitation frequency, but moves on to resonance for a $\pi/2$ pulse, moves to a frequency above the excitation frequency for a $\pi$ pulse, and then returns to resonance for a $3\pi/2$ pulse. Rabi oscillations performed in this way achieve nearly complete magnetization inversion, as evidenced by the minimum being almost zero in Figure \ref{fig:RabiCurveFigure}c.

\begin{figure}
    \centering
    \includegraphics{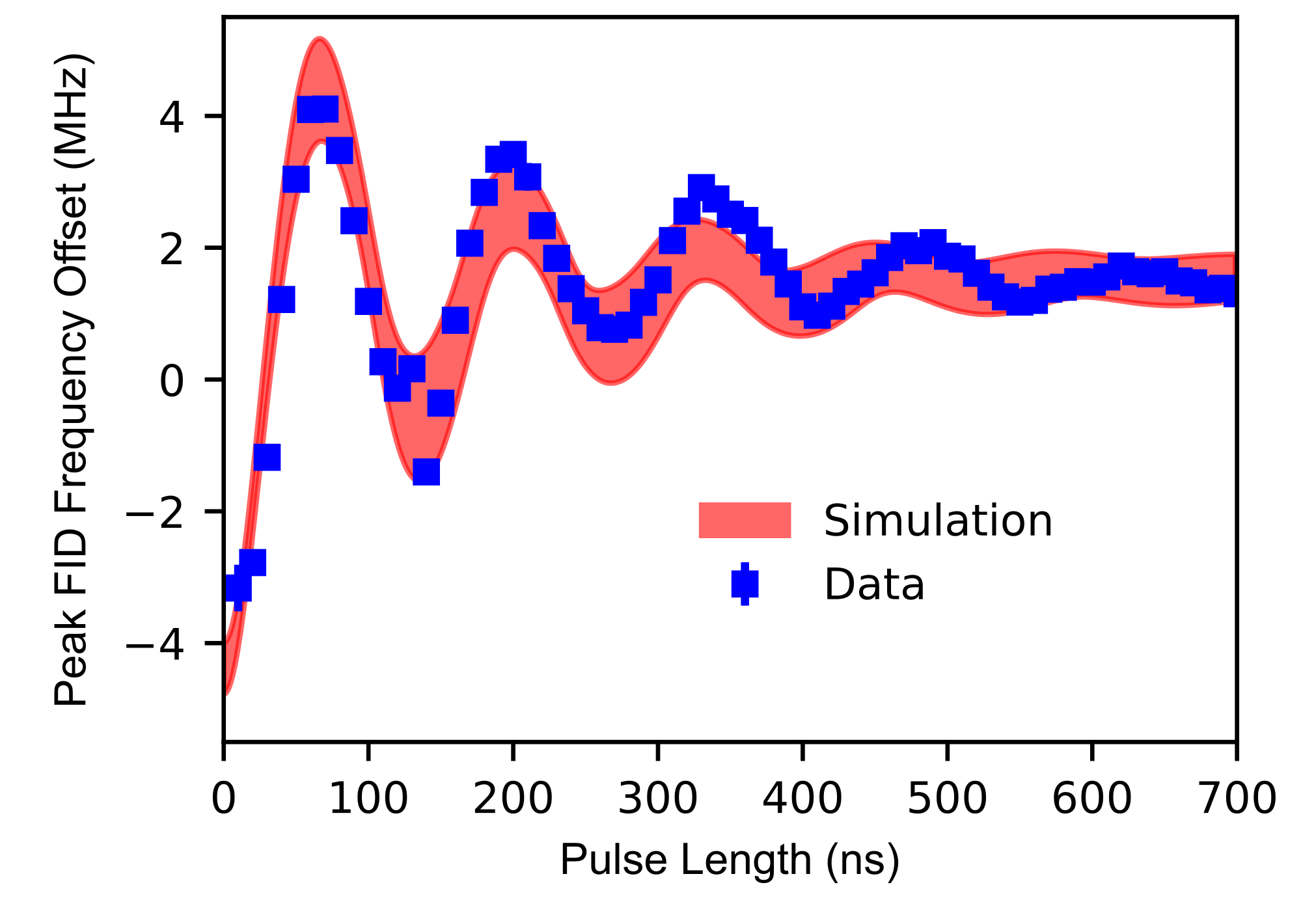}
    \caption{
    Mean FID frequency as a function of pulse length as BDPA-Bz crystal undergoes Rabi oscillations over many cycles. Simulated mean FID frequencies match experimental results over multiple oscillations.
    }
    \label{fig:LongPulseFigure}
\end{figure}

Numerical simulations of the nonlinear modified Bloch equations with excitation pulses of varying lengths were carried out assuming an oblate spheroid geometry with an aspect ratio of 7:1, for an initial magnetization $M_0 = 270$ A/m. The phenomenological Bloch relaxation times $T_1$ and $T_2$ were independently measured to be $270 \pm 40$ ns and $120 \pm 30$ ns, respectively at 8.56 T (see Supplemental Material \cite{SI}), in good agreement with values measured at low field \cite{Mitchell_Electron_2011}. Simulations of the FID frequency dependence on pulse length over a full Rabi oscillation were in excellent agreement with experimental results, as shown in Figure \ref{fig:RabiCurveFigure}d. The width of the simulated frequency shift indicates a 95\% confidence interval, taking into account uncertainties in $T_1$, $T_2$, the Rabi frequency $\omega_1$, and the FEL detuning (see Supplemental Material \cite{SI}).

Magnetization evolution was probed over many Rabi cycles by varying the pulse length from 0 - 700 ns in steps of 10 ns, using an on-resonance Rabi frequency $\omega_1/2\pi = 8$ MHz (Figure \ref{fig:LongPulseFigure}). At least five Rabi cycles were observed. The width of the simulation indicates a 95\% confidence interval, and is in excellent agreement with experimental results. The decay in the observed Rabi oscillations is primarily related to spin-lattice ($T_1$) and spin-spin ($T_2$) relaxation.

\begin{figure}
    \centering
    \includegraphics{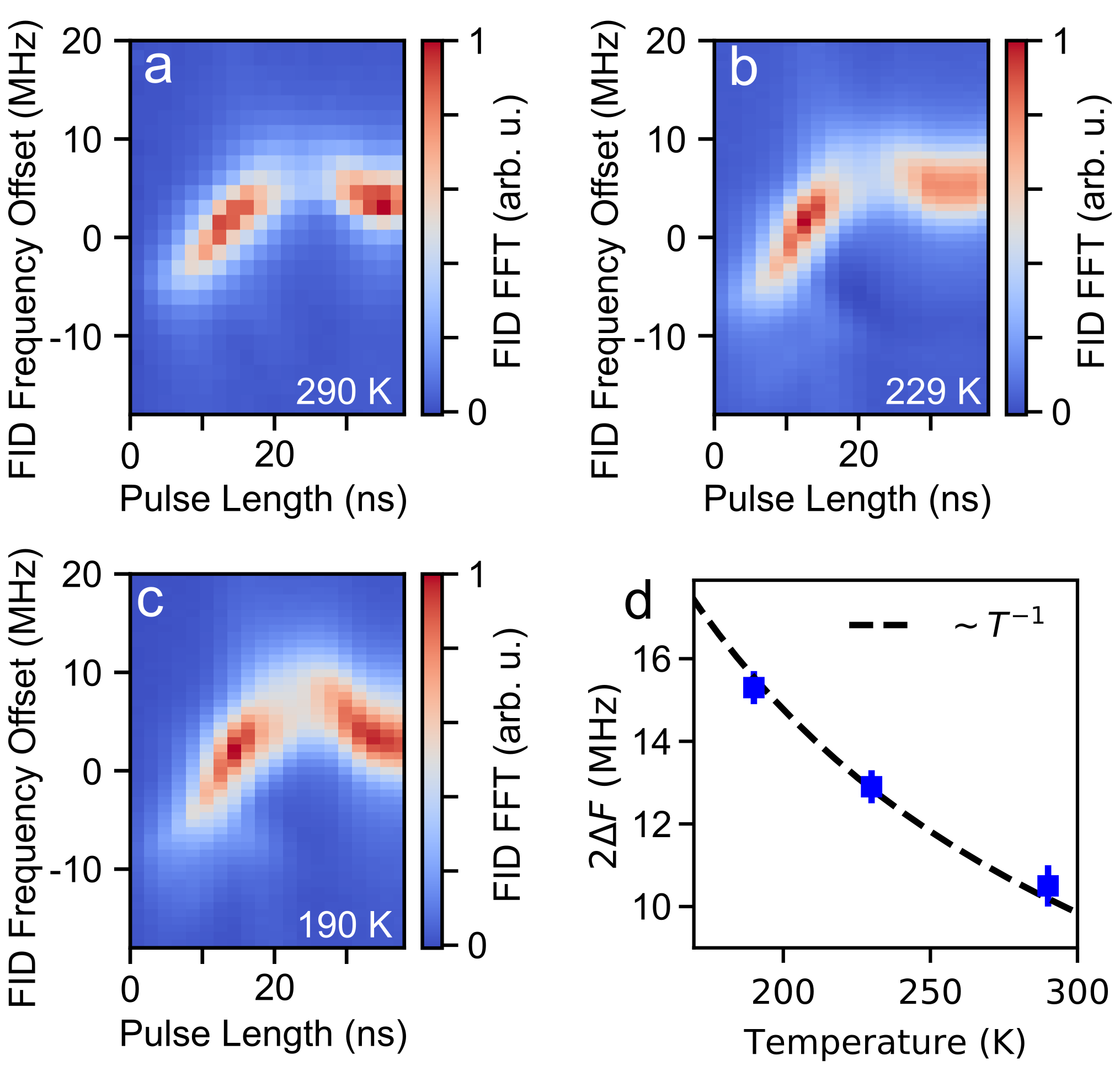}
    \caption{
    Contour plots of the FT-EPR lineshape at \textbf{a} 290 K, \textbf{b} 229 K, and \textbf{c} 190 K. \textbf{d} The maximum frequency shift $\Delta F$ as a function of temperature.
    }
    \label{fig:TemperatureFigure}
\end{figure}

The first condition for resolving a tip-angle dependent frequency shift in the dressed Rabi oscillation phase is that the ratio $|\Delta \omega/\Gamma|\geq$ 1, where $\Delta\omega = -\mu_0\gamma\theta_d M_0$ is the demagnetizing field-induced frequency shift and $\Gamma$ is the EPR linewidth. For BDPA-Bz grains at 8.56 T, $\Gamma \simeq$ 4 - 6 MHz. For $\theta_d = 0.65$ used in simulations described above, we have $|\Delta\omega/\Gamma| \simeq 1.3 $. The second condition is that the ratio $ \omega_1/\Gamma\geq$ 1 , where $ \omega_1$ is the Rabi frequency. For all previous EPR studies we are aware of, the combination of a sufficiently large equilibrium magnetization, narrow linewidth, and large Rabi frequency has been difficult to achieve, $|\Delta\omega/\Gamma| \ll 1$, and resolving magnetization-dependent frequency shifts has been difficult or impossible.

The temperature dependence of the fractional frequency shift $|\Delta \omega/\omega_L|$ in the dressed Rabi oscillation phase can be estimated for a paramagnet consisting of spin-1/2 electrons with density $n$ at thermal equilibrium, where $M_0 = \frac{g\mu_B}{2} n\tanh \frac{\mu_0\gamma H_0}{2 k_B T}$. When the high temperature approximation $k_B T \gg \mu_0 \gamma \hbar H_0$ is valid,
\begin{align}
    \frac{\Delta \omega}{\omega_L} = \mu_0 n \frac{(\gamma\hbar)^2}{4 k_B T}
    \label{eq:fractional_freq_shift}
\end{align}
Figure \ref{fig:TemperatureFigure} shows contour plots of nutation experiments performed at room temperature, 229 K, and 190 K. The temperature dependence of the maximum observed frequency shift $\Delta F$ is consistent with the predicted temperature dependence (Figure \ref{fig:TemperatureFigure}a). However, as the temperature is lowered, evidence of more complicated dynamics emerges in the nutation experiment: small additional frequency components are observed in FIDs for pulse durations shorter than 20 ns, as shown in Figures \ref{fig:TemperatureFigure}b and \ref{fig:TemperatureFigure}c.


Demagnetization field effects are commonly encountered in ferromagnetic resonance \cite{Kittel_Ferro_1948}, where sample geometry plays an important role in determining the resonance condition at very small microwave fields and tip angles. However, as microwave fields are increased, even in spherical samples, spin wave instabilities destroy the state with spatially-uniform magnetization before significant flip angles can be achieved \cite{Suhl_1956}. Demagnetization field effects are also encountered in nuclear magnetic resonance (NMR) in highly concentrated spin systems \cite{Levitt_Demagnetization_1996}. Nonlinear magnetization-dependent effects in NMR can manifest in the generation of multiple spin-echos by a pair of radio frequency pulses \cite{Bowtell_Multiple_1969, Denville_NMR_1979, Bedford_Multiple_1991} or in anomalous frequency correlations appearing in correlation spectroscopy experiments \cite{Warren_Generation_1993, Richter_Imaging_1995, Jeener_Equivalence_2000}. Taking $\theta_d = 1$, corresponding to a flat disk geometry, Equation \ref{eq:fractional_freq_shift} predicts the $^1$H NMR resonance measured in room-temperature water should shift by $\sim4$ ppb, corresponding to a frequency shift of 1.6 Hz in a 9.4 T magnetic field. In 1990, Edzes reported an anomalous $^1$H NMR frequency shift of $\sim 2.5$ ppb in protonated solvents, consistent with Equation \ref{eq:fractional_freq_shift} for $\theta_d \sim 0.6$ \cite{Edzes_TheNuclear_1969}. Outside of this example, demagnetization field-induced frequency shifts are not commonly observed in NMR, which can be explained by the fact that the gyromagnetic ratio of electrons is 657 times larger than that of protons, together with the $\gamma^2$ dependence of Equation \ref{eq:fractional_freq_shift}.

Another nonlinear, magnetization-dependent effect encountered in magnetic resonance experiments is radiation damping \cite{Bloembergen_Radiation_1954}. Radiation damping typically manifests as magnetization and tip-angle dependent line broadening \cite{Augustine_Transient_2002} in magnetic resonance experiments which employ resonators with $Q \gg 1$, with high filling factors \cite{Benner_Influence_977,Prisner_Pulsed_1992}. The effects of radiation damping are small in our experiments, since we do not employ a resonator cavity and our effective filling factor $\eta \ll 1$, but may be responsible for some line broadening (see Supplemental Material \cite{SI}).

Direct radiative losses due to magnetic dipole radiation may also contribute to the observed EPR linewidths, especially at low temperatures. The instantaneous power $P$ radiated by magnetic dipole radiation at frequency $\omega$ by a sample of volume $V$ with uniform magnetization $M_0$ is given by $P = \frac{\mu_0}{6\pi c^3} \omega^4 M_0^2 V^2 \sin^2\theta$, where $\theta$ is the magnetization tip-angle. Radiative losses lead to magnetization decay with characteristic time constant $\tau_{rad} = (\frac{\mu_0}{6\pi c^3}\gamma \omega^3 M_0 V)^{-1}$ (see Supplemental Material \cite{SI}). For the grain used in experiments presented in Figure \ref{fig:RabiCurveFigure}, a lower limit of $\tau_{rad} \simeq 360$ ns was calculated, which is longer than $T_2$, $T_2^* = 1/\pi\Gamma$, or $T_1$, but which may become important to the linewidth at low temperatures.

Nonlinear spin-spin interactions like those reported in this Letter are essential for developing the quantum-mechanical correlations between spins necessary to generate squeezed spin states \cite{Kitagawa_Squeezed_1993}. Schemes for realizing squeezed states of electron spin ensembles have been proposed involving phonon-induced interactions between nitrogen-vacancy (NV) centers in diamond \cite{Bennet_Phonon_2013}, or by coupling a NV center ensemble to a magnetic tip attached to a nanomechanical resonator \cite{Ma_Bistability_2016}. Demagnetization field effects offer an alternative method of engineering nonlinear spin-spin interactions. The nonlinear interaction strength can be tuned by adjusting the magnetization magnitude through temperature changes, as shown in Figure \ref{fig:RabiCurveFigure}f, as well as by optimizing the sample geometry, as illustrated by Equation \ref{eq:BlochSpelledOut}. The long-range magnetic interactions that dress Rabi oscillations in this work may also be useful for coupling distant spin qubits \cite{Trifunovic_2013}.

In this Letter, we have shown that, when both the microwave field and the demagnetization field are larger than the line width of a magnetic resonance, the non-interacting spin dynamics may be visibly ``dressed” by each spin’s interaction with the mean field generated by all the other spins in the sample. The signatures of dressed Rabi oscillations are free-induction decays whose frequencies depend on tip angle, and anomalous non-circular trajectories of the sample magnetization direction on the surface of the Bloch sphere. Because the experiments are carried out without a resonator, the ``radiation damping” effects that usually arise in the spin dynamics of concentrated spin systems are negligible. We think of dressed Rabi oscillation as a transient dynamical phase in a system of interacting spins driven far from thermal equilibrium. Dressed Rabi oscillation can be modeled using classical equations of motion. As the temperature is lowered, in addition to the strength of the ``dressing” increasing, the thermal fluctuations in the exchange fields that are responsible for narrowing the EPR line near room temperature will be quenched, and signatures of new manifestly quantum states of interacting spins far from thermal equilibrium may emerge.


\begin{acknowledgments}
The authors acknowledge David Enyeart and Nickolay Agladze for maintaining, repairing, and assisting with operation of the UCSB FEL, and Gerald Ramian for many useful discussions. Support for this work came from the National Science Foundation through NSF-MCB-1617025. This work was performed at the ITST Terahertz Facilities at UCSB, which have been upgraded under NSF Awards DMR-1126894 and DMR-1626681.
\end{acknowledgments}


\section{References}
\label{References}

\bibliography{biblio}

\begin{thebibliography}{50}%
\makeatletter
\providecommand \@ifxundefined [1]{%
 \@ifx{#1\undefined}
}%
\providecommand \@ifnum [1]{%
 \ifnum #1\expandafter \@firstoftwo
 \else \expandafter \@secondoftwo
 \fi
}%
\providecommand \@ifx [1]{%
 \ifx #1\expandafter \@firstoftwo
 \else \expandafter \@secondoftwo
 \fi
}%
\providecommand \natexlab [1]{#1}%
\providecommand \enquote  [1]{``#1''}%
\providecommand \bibnamefont  [1]{#1}%
\providecommand \bibfnamefont [1]{#1}%
\providecommand \citenamefont [1]{#1}%
\providecommand \href@noop [0]{\@secondoftwo}%
\providecommand \href [0]{\begingroup \@sanitize@url \@href}%
\providecommand \@href[1]{\@@startlink{#1}\@@href}%
\providecommand \@@href[1]{\endgroup#1\@@endlink}%
\providecommand \@sanitize@url [0]{\catcode `\\12\catcode `\$12\catcode
  `\&12\catcode `\#12\catcode `\^12\catcode `\_12\catcode `\%12\relax}%
\providecommand \@@startlink[1]{}%
\providecommand \@@endlink[0]{}%
\providecommand \url  [0]{\begingroup\@sanitize@url \@url }%
\providecommand \@url [1]{\endgroup\@href {#1}{\urlprefix }}%
\providecommand \urlprefix  [0]{URL }%
\providecommand \Eprint [0]{\href }%
\providecommand \doibase [0]{http://dx.doi.org/}%
\providecommand \selectlanguage [0]{\@gobble}%
\providecommand \bibinfo  [0]{\@secondoftwo}%
\providecommand \bibfield  [0]{\@secondoftwo}%
\providecommand \translation [1]{[#1]}%
\providecommand \BibitemOpen [0]{}%
\providecommand \bibitemStop [0]{}%
\providecommand \bibitemNoStop [0]{.\EOS\space}%
\providecommand \EOS [0]{\spacefactor3000\relax}%
\providecommand \BibitemShut  [1]{\csname bibitem#1\endcsname}%
\let\auto@bib@innerbib\@empty
\bibitem [{\citenamefont {Anderson}(1973)}]{Anderson_Resonating_1973}%
  \BibitemOpen
  \bibfield  {author} {\bibinfo {author} {\bibfnamefont {P.}~\bibnamefont
  {Anderson}},\ }\href {\doibase https://doi.org/10.1016/0025-5408(73)90167-0}
  {\bibfield  {journal} {\bibinfo  {journal} {Materials Research Bulletin}\
  }\textbf {\bibinfo {volume} {8}},\ \bibinfo {pages} {153 } (\bibinfo {year}
  {1973})}\BibitemShut {NoStop}%
\bibitem [{\citenamefont {Senthil}\ and\ \citenamefont
  {Fisher}(2000)}]{Senthil_Z2_2000}%
  \BibitemOpen
  \bibfield  {author} {\bibinfo {author} {\bibfnamefont {T.}~\bibnamefont
  {Senthil}}\ and\ \bibinfo {author} {\bibfnamefont {M.~P.~A.}\ \bibnamefont
  {Fisher}},\ }\href {\doibase 10.1103/PhysRevB.62.7850} {\bibfield  {journal}
  {\bibinfo  {journal} {Phys. Rev. B}\ }\textbf {\bibinfo {volume} {62}},\
  \bibinfo {pages} {7850} (\bibinfo {year} {2000})}\BibitemShut {NoStop}%
\bibitem [{\citenamefont {Moessner}\ and\ \citenamefont
  {Sondhi}(2001)}]{Moessner_Resonating_2001}%
  \BibitemOpen
  \bibfield  {author} {\bibinfo {author} {\bibfnamefont {R.}~\bibnamefont
  {Moessner}}\ and\ \bibinfo {author} {\bibfnamefont {S.~L.}\ \bibnamefont
  {Sondhi}},\ }\href {\doibase 10.1103/PhysRevLett.86.1881} {\bibfield
  {journal} {\bibinfo  {journal} {Phys. Rev. Lett.}\ }\textbf {\bibinfo
  {volume} {86}},\ \bibinfo {pages} {1881} (\bibinfo {year}
  {2001})}\BibitemShut {NoStop}%
\bibitem [{\citenamefont {Kitaev}(2006)}]{Kitaev_Anyons_2006}%
  \BibitemOpen
  \bibfield  {author} {\bibinfo {author} {\bibfnamefont {A.}~\bibnamefont
  {Kitaev}},\ }\href {\doibase https://doi.org/10.1016/j.aop.2005.10.005}
  {\bibfield  {journal} {\bibinfo  {journal} {Annals of Physics}\ }\textbf
  {\bibinfo {volume} {321}},\ \bibinfo {pages} {2 } (\bibinfo {year} {2006})},\
  \bibinfo {note} {january Special Issue}\BibitemShut {NoStop}%
\bibitem [{\citenamefont {Balents}(2010)}]{balents2010spin}%
  \BibitemOpen
  \bibfield  {author} {\bibinfo {author} {\bibfnamefont {L.}~\bibnamefont
  {Balents}},\ }\href@noop {} {\bibfield  {journal} {\bibinfo  {journal}
  {Nature}\ }\textbf {\bibinfo {volume} {464}},\ \bibinfo {pages} {199}
  (\bibinfo {year} {2010})}\BibitemShut {NoStop}%
\bibitem [{\citenamefont {Powell}\ and\ \citenamefont
  {McKenzie}(2011)}]{Powell_Quantum_2011}%
  \BibitemOpen
  \bibfield  {author} {\bibinfo {author} {\bibfnamefont {B.~J.}\ \bibnamefont
  {Powell}}\ and\ \bibinfo {author} {\bibfnamefont {R.~H.}\ \bibnamefont
  {McKenzie}},\ }\href {\doibase 10.1088/0034-4885/74/5/056501} {\bibfield
  {journal} {\bibinfo  {journal} {Reports on Progress in Physics}\ }\textbf
  {\bibinfo {volume} {74}},\ \bibinfo {pages} {056501} (\bibinfo {year}
  {2011})}\BibitemShut {NoStop}%
\bibitem [{\citenamefont {Savary}\ and\ \citenamefont
  {Balents}(2016)}]{Savary_2016}%
  \BibitemOpen
  \bibfield  {author} {\bibinfo {author} {\bibfnamefont {L.}~\bibnamefont
  {Savary}}\ and\ \bibinfo {author} {\bibfnamefont {L.}~\bibnamefont
  {Balents}},\ }\href {\doibase 10.1088/0034-4885/80/1/016502} {\bibfield
  {journal} {\bibinfo  {journal} {Reports on Progress in Physics}\ }\textbf
  {\bibinfo {volume} {80}},\ \bibinfo {pages} {016502} (\bibinfo {year}
  {2016})}\BibitemShut {NoStop}%
\bibitem [{\citenamefont {Vojta}(2018)}]{Vojta_2018}%
  \BibitemOpen
  \bibfield  {author} {\bibinfo {author} {\bibfnamefont {M.}~\bibnamefont
  {Vojta}},\ }\href {\doibase 10.1088/1361-6633/aab6be} {\bibfield  {journal}
  {\bibinfo  {journal} {Reports on Progress in Physics}\ }\textbf {\bibinfo
  {volume} {81}},\ \bibinfo {pages} {064501} (\bibinfo {year}
  {2018})}\BibitemShut {NoStop}%
\bibitem [{\citenamefont {Else}\ \emph {et~al.}(2016)\citenamefont {Else},
  \citenamefont {Bauer},\ and\ \citenamefont {Nayak}}]{Floquet_time_crystals}%
  \BibitemOpen
  \bibfield  {author} {\bibinfo {author} {\bibfnamefont {D.~V.}\ \bibnamefont
  {Else}}, \bibinfo {author} {\bibfnamefont {B.}~\bibnamefont {Bauer}}, \ and\
  \bibinfo {author} {\bibfnamefont {C.}~\bibnamefont {Nayak}},\ }\href
  {\doibase 10.1103/PhysRevLett.117.090402} {\bibfield  {journal} {\bibinfo
  {journal} {Phys. Rev. Lett.}\ }\textbf {\bibinfo {volume} {117}},\ \bibinfo
  {pages} {090402} (\bibinfo {year} {2016})}\BibitemShut {NoStop}%
\bibitem [{\citenamefont {Choi}\ \emph {et~al.}(2017)\citenamefont {Choi},
  \citenamefont {Choi}, \citenamefont {Landig}, \citenamefont {Kucsko},
  \citenamefont {Zhou}, \citenamefont {Isoya}, \citenamefont {Jelezko},
  \citenamefont {Onoda}, \citenamefont {Sumiya}, \citenamefont {Khemani} \emph
  {et~al.}}]{choi2017observation}%
  \BibitemOpen
  \bibfield  {author} {\bibinfo {author} {\bibfnamefont {S.}~\bibnamefont
  {Choi}}, \bibinfo {author} {\bibfnamefont {J.}~\bibnamefont {Choi}}, \bibinfo
  {author} {\bibfnamefont {R.}~\bibnamefont {Landig}}, \bibinfo {author}
  {\bibfnamefont {G.}~\bibnamefont {Kucsko}}, \bibinfo {author} {\bibfnamefont
  {H.}~\bibnamefont {Zhou}}, \bibinfo {author} {\bibfnamefont {J.}~\bibnamefont
  {Isoya}}, \bibinfo {author} {\bibfnamefont {F.}~\bibnamefont {Jelezko}},
  \bibinfo {author} {\bibfnamefont {S.}~\bibnamefont {Onoda}}, \bibinfo
  {author} {\bibfnamefont {H.}~\bibnamefont {Sumiya}}, \bibinfo {author}
  {\bibfnamefont {V.}~\bibnamefont {Khemani}},  \emph {et~al.},\ }\href@noop {}
  {\bibfield  {journal} {\bibinfo  {journal} {Nature}\ }\textbf {\bibinfo
  {volume} {543}},\ \bibinfo {pages} {221} (\bibinfo {year}
  {2017})}\BibitemShut {NoStop}%
\bibitem [{\citenamefont {Duffy}\ \emph {et~al.}(1972)\citenamefont {Duffy},
  \citenamefont {Dubach}, \citenamefont {Pianetta}, \citenamefont {Deck},
  \citenamefont {Strandburg},\ and\ \citenamefont
  {Miedema}}]{Duffy_Antiferromagnetic_1972}%
  \BibitemOpen
  \bibfield  {author} {\bibinfo {author} {\bibfnamefont {W.}~\bibnamefont
  {Duffy}}, \bibinfo {author} {\bibfnamefont {J.~F.}\ \bibnamefont {Dubach}},
  \bibinfo {author} {\bibfnamefont {P.~A.}\ \bibnamefont {Pianetta}}, \bibinfo
  {author} {\bibfnamefont {J.~F.}\ \bibnamefont {Deck}}, \bibinfo {author}
  {\bibfnamefont {D.~L.}\ \bibnamefont {Strandburg}}, \ and\ \bibinfo {author}
  {\bibfnamefont {A.~R.}\ \bibnamefont {Miedema}},\ }\href {\doibase
  doi:http://dx.doi.org/10.1063/1.1677580} {\bibfield  {journal} {\bibinfo
  {journal} {The Journal of Chemical Physics}\ }\textbf {\bibinfo {volume}
  {56}},\ \bibinfo {pages} {2555} (\bibinfo {year} {1972})}\BibitemShut
  {NoStop}%
\bibitem [{\citenamefont {Azuma}\ \emph {et~al.}(1994)\citenamefont {Azuma},
  \citenamefont {Ozawa},\ and\ \citenamefont
  {Yamauchi}}]{Azuma_Molecular_1994}%
  \BibitemOpen
  \bibfield  {author} {\bibinfo {author} {\bibfnamefont {N.}~\bibnamefont
  {Azuma}}, \bibinfo {author} {\bibfnamefont {T.}~\bibnamefont {Ozawa}}, \ and\
  \bibinfo {author} {\bibfnamefont {J.}~\bibnamefont {Yamauchi}},\ }\href
  {\doibase 10.1246/bcsj.67.31} {\bibfield  {journal} {\bibinfo  {journal}
  {Bulletin of the Chemical Society of Japan}\ }\textbf {\bibinfo {volume}
  {67}},\ \bibinfo {pages} {31} (\bibinfo {year} {1994})}\BibitemShut {NoStop}%
\bibitem [{\citenamefont {Mitchell}\ \emph {et~al.}(2011)\citenamefont
  {Mitchell}, \citenamefont {Quine}, \citenamefont {Tseitlin}, \citenamefont
  {Weber}, \citenamefont {Meyer}, \citenamefont {Avery}, \citenamefont
  {Eaton},\ and\ \citenamefont {Eaton}}]{Mitchell_Electron_2011}%
  \BibitemOpen
  \bibfield  {author} {\bibinfo {author} {\bibfnamefont {D.~G.}\ \bibnamefont
  {Mitchell}}, \bibinfo {author} {\bibfnamefont {R.~W.}\ \bibnamefont {Quine}},
  \bibinfo {author} {\bibfnamefont {M.}~\bibnamefont {Tseitlin}}, \bibinfo
  {author} {\bibfnamefont {R.~T.}\ \bibnamefont {Weber}}, \bibinfo {author}
  {\bibfnamefont {V.}~\bibnamefont {Meyer}}, \bibinfo {author} {\bibfnamefont
  {A.}~\bibnamefont {Avery}}, \bibinfo {author} {\bibfnamefont {S.~S.}\
  \bibnamefont {Eaton}}, \ and\ \bibinfo {author} {\bibfnamefont {G.~R.}\
  \bibnamefont {Eaton}},\ }\href {\doibase 10.1021/jp201978w} {\bibfield
  {journal} {\bibinfo  {journal} {The Journal of Physical Chemistry B}\
  }\textbf {\bibinfo {volume} {115}},\ \bibinfo {pages} {7986} (\bibinfo {year}
  {2011})},\ \bibinfo {note} {pMID: 21574594},\ \Eprint
  {http://arxiv.org/abs/https://doi.org/10.1021/jp201978w}
  {https://doi.org/10.1021/jp201978w} \BibitemShut {NoStop}%
\bibitem [{\citenamefont {Yamauchi}\ and\ \citenamefont
  {Deguchi}(1977)}]{Yamauchi_Low_1977}%
  \BibitemOpen
  \bibfield  {author} {\bibinfo {author} {\bibfnamefont {J.}~\bibnamefont
  {Yamauchi}}\ and\ \bibinfo {author} {\bibfnamefont {Y.}~\bibnamefont
  {Deguchi}},\ }\href {\doibase 10.1246/bcsj.50.2803} {\bibfield  {journal}
  {\bibinfo  {journal} {Bulletin of the Chemical Society of Japan}\ }\textbf
  {\bibinfo {volume} {50}},\ \bibinfo {pages} {2803} (\bibinfo {year}
  {1977})}\BibitemShut {NoStop}%
\bibitem [{\citenamefont {Hamilton}\ and\ \citenamefont
  {Pake}(1963)}]{Hamilton_Linear_1963}%
  \BibitemOpen
  \bibfield  {author} {\bibinfo {author} {\bibfnamefont {W.~O.}\ \bibnamefont
  {Hamilton}}\ and\ \bibinfo {author} {\bibfnamefont {G.~E.}\ \bibnamefont
  {Pake}},\ }\href {\doibase 10.1063/1.1734084} {\bibfield  {journal} {\bibinfo
   {journal} {The Journal of Chemical Physics}\ }\textbf {\bibinfo {volume}
  {39}},\ \bibinfo {pages} {2694} (\bibinfo {year} {1963})},\ \Eprint
  {http://arxiv.org/abs/https://doi.org/10.1063/1.1734084}
  {https://doi.org/10.1063/1.1734084} \BibitemShut {NoStop}%
\bibitem [{\citenamefont {Bennati}\ \emph {et~al.}(1999)\citenamefont
  {Bennati}, \citenamefont {Farrar}, \citenamefont {Bryant}, \citenamefont
  {Inati}, \citenamefont {Weis}, \citenamefont {Gerfen}, \citenamefont
  {Riggs-Gelasco}, \citenamefont {Stubbe},\ and\ \citenamefont
  {Griffin}}]{Bennati_Pulsed_1999}%
  \BibitemOpen
  \bibfield  {author} {\bibinfo {author} {\bibfnamefont {M.}~\bibnamefont
  {Bennati}}, \bibinfo {author} {\bibfnamefont {C.}~\bibnamefont {Farrar}},
  \bibinfo {author} {\bibfnamefont {J.}~\bibnamefont {Bryant}}, \bibinfo
  {author} {\bibfnamefont {S.}~\bibnamefont {Inati}}, \bibinfo {author}
  {\bibfnamefont {V.}~\bibnamefont {Weis}}, \bibinfo {author} {\bibfnamefont
  {G.}~\bibnamefont {Gerfen}}, \bibinfo {author} {\bibfnamefont
  {P.}~\bibnamefont {Riggs-Gelasco}}, \bibinfo {author} {\bibfnamefont
  {J.}~\bibnamefont {Stubbe}}, \ and\ \bibinfo {author} {\bibfnamefont
  {R.}~\bibnamefont {Griffin}},\ }\href {\doibase
  https://doi.org/10.1006/jmre.1999.1727} {\bibfield  {journal} {\bibinfo
  {journal} {Journal of Magnetic Resonance}\ }\textbf {\bibinfo {volume}
  {138}},\ \bibinfo {pages} {232 } (\bibinfo {year} {1999})}\BibitemShut
  {NoStop}%
\bibitem [{\citenamefont {Goldfarb}\ \emph {et~al.}(2008)\citenamefont
  {Goldfarb}, \citenamefont {Lipkin}, \citenamefont {Potapov}, \citenamefont
  {Gorodetsky}, \citenamefont {Epel}, \citenamefont {Raitsimring},
  \citenamefont {Radoul},\ and\ \citenamefont
  {Kaminker}}]{Goldfarb_HYSCORE_2008}%
  \BibitemOpen
  \bibfield  {author} {\bibinfo {author} {\bibfnamefont {D.}~\bibnamefont
  {Goldfarb}}, \bibinfo {author} {\bibfnamefont {Y.}~\bibnamefont {Lipkin}},
  \bibinfo {author} {\bibfnamefont {A.}~\bibnamefont {Potapov}}, \bibinfo
  {author} {\bibfnamefont {Y.}~\bibnamefont {Gorodetsky}}, \bibinfo {author}
  {\bibfnamefont {B.}~\bibnamefont {Epel}}, \bibinfo {author} {\bibfnamefont
  {A.~M.}\ \bibnamefont {Raitsimring}}, \bibinfo {author} {\bibfnamefont
  {M.}~\bibnamefont {Radoul}}, \ and\ \bibinfo {author} {\bibfnamefont
  {I.}~\bibnamefont {Kaminker}},\ }\href {\doibase
  https://doi.org/10.1016/j.jmr.2008.05.019} {\bibfield  {journal} {\bibinfo
  {journal} {Journal of Magnetic Resonance}\ }\textbf {\bibinfo {volume}
  {194}},\ \bibinfo {pages} {8 } (\bibinfo {year} {2008})}\BibitemShut
  {NoStop}%
\bibitem [{\citenamefont {Durkan}\ and\ \citenamefont
  {Welland}(2002)}]{Durkan_Electronic_2002}%
  \BibitemOpen
  \bibfield  {author} {\bibinfo {author} {\bibfnamefont {C.}~\bibnamefont
  {Durkan}}\ and\ \bibinfo {author} {\bibfnamefont {M.~E.}\ \bibnamefont
  {Welland}},\ }\href {\doibase 10.1063/1.1434301} {\bibfield  {journal}
  {\bibinfo  {journal} {Applied Physics Letters}\ }\textbf {\bibinfo {volume}
  {80}},\ \bibinfo {pages} {458} (\bibinfo {year} {2002})},\ \Eprint
  {http://arxiv.org/abs/https://doi.org/10.1063/1.1434301}
  {https://doi.org/10.1063/1.1434301} \BibitemShut {NoStop}%
\bibitem [{\citenamefont {Dane}\ and\ \citenamefont
  {Swager}(2010)}]{Dane_Synthesis_2010}%
  \BibitemOpen
  \bibfield  {author} {\bibinfo {author} {\bibfnamefont {E.~L.}\ \bibnamefont
  {Dane}}\ and\ \bibinfo {author} {\bibfnamefont {T.~M.}\ \bibnamefont
  {Swager}},\ }\href {\doibase 10.1021/jo100577g} {\bibfield  {journal}
  {\bibinfo  {journal} {The Journal of Organic Chemistry}\ }\textbf {\bibinfo
  {volume} {75}},\ \bibinfo {pages} {3533} (\bibinfo {year} {2010})},\ \Eprint
  {http://arxiv.org/abs/https://doi.org/10.1021/jo100577g}
  {https://doi.org/10.1021/jo100577g} \BibitemShut {NoStop}%
\bibitem [{\citenamefont {Giraudeau}\ \emph {et~al.}(2009)\citenamefont
  {Giraudeau}, \citenamefont {Shrot},\ and\ \citenamefont
  {Frydman}}]{Giraudeau_Multiple_2009}%
  \BibitemOpen
  \bibfield  {author} {\bibinfo {author} {\bibfnamefont {P.}~\bibnamefont
  {Giraudeau}}, \bibinfo {author} {\bibfnamefont {Y.}~\bibnamefont {Shrot}}, \
  and\ \bibinfo {author} {\bibfnamefont {L.}~\bibnamefont {Frydman}},\ }\href
  {\doibase 10.1021/ja905096f} {\bibfield  {journal} {\bibinfo  {journal}
  {Journal of the American Chemical Society}\ }\textbf {\bibinfo {volume}
  {131}},\ \bibinfo {pages} {13902} (\bibinfo {year} {2009})},\ \bibinfo {note}
  {pMID: 19743849},\ \Eprint
  {http://arxiv.org/abs/https://doi.org/10.1021/ja905096f}
  {https://doi.org/10.1021/ja905096f} \BibitemShut {NoStop}%
\bibitem [{\citenamefont {Anderson}\ and\ \citenamefont
  {Weiss}(1953)}]{Anderson_Exchange_1953}%
  \BibitemOpen
  \bibfield  {author} {\bibinfo {author} {\bibfnamefont {P.~W.}\ \bibnamefont
  {Anderson}}\ and\ \bibinfo {author} {\bibfnamefont {P.~R.}\ \bibnamefont
  {Weiss}},\ }\href {\doibase 10.1103/RevModPhys.25.269} {\bibfield  {journal}
  {\bibinfo  {journal} {Rev. Mod. Phys.}\ }\textbf {\bibinfo {volume} {25}},\
  \bibinfo {pages} {269} (\bibinfo {year} {1953})}\BibitemShut {NoStop}%
\bibitem [{\citenamefont {Misra}(2011)}]{Sushil_Multifrequency_2011}%
  \BibitemOpen
  \bibfield  {author} {\bibinfo {author} {\bibfnamefont {S.~K.}\ \bibnamefont
  {Misra}},\ }\href@noop {} {\emph {\bibinfo {title} {Multifrequency Electron
  Paramagnetic Resonance : Theory and Applications}}}\ (\bibinfo  {publisher}
  {John Wiley and Sons, Incorporated},\ \bibinfo {address} {Hoboken},\ \bibinfo
  {year} {2011})\BibitemShut {NoStop}%
\bibitem [{\citenamefont {Hegmann}\ and\ \citenamefont
  {Sherwin}(1996)}]{hegmann_generation_1996}%
  \BibitemOpen
  \bibfield  {author} {\bibinfo {author} {\bibfnamefont {F.~A.}\ \bibnamefont
  {Hegmann}}\ and\ \bibinfo {author} {\bibfnamefont {M.~S.}\ \bibnamefont
  {Sherwin}}\ }(\bibinfo  {publisher} {International Society for Optics and
  Photonics},\ \bibinfo {year} {1996})\ pp.\ \bibinfo {pages}
  {90--106}\BibitemShut {NoStop}%
\bibitem [{\citenamefont {Takahashi}\ \emph {et~al.}(2009)\citenamefont
  {Takahashi}, \citenamefont {Ramian},\ and\ \citenamefont
  {Sherwin}}]{Takahashi_Cavity_2009}%
  \BibitemOpen
  \bibfield  {author} {\bibinfo {author} {\bibfnamefont {S.}~\bibnamefont
  {Takahashi}}, \bibinfo {author} {\bibfnamefont {G.}~\bibnamefont {Ramian}}, \
  and\ \bibinfo {author} {\bibfnamefont {M.~S.}\ \bibnamefont {Sherwin}},\
  }\href {\doibase 10.1063/1.3270041} {\bibfield  {journal} {\bibinfo
  {journal} {Applied Physics Letters}\ }\textbf {\bibinfo {volume} {95}},\
  \bibinfo {pages} {3} (\bibinfo {year} {2009})}\BibitemShut {NoStop}%
\bibitem [{\citenamefont {Takahashi}\ \emph {et~al.}(2007)\citenamefont
  {Takahashi}, \citenamefont {Ramian}, \citenamefont {Sherwin}, \citenamefont
  {Brunel},\ and\ \citenamefont {van Tol}}]{Takahashi2007}%
  \BibitemOpen
  \bibfield  {author} {\bibinfo {author} {\bibfnamefont {S.}~\bibnamefont
  {Takahashi}}, \bibinfo {author} {\bibfnamefont {G.}~\bibnamefont {Ramian}},
  \bibinfo {author} {\bibfnamefont {M.~S.}\ \bibnamefont {Sherwin}}, \bibinfo
  {author} {\bibfnamefont {L.-C.}\ \bibnamefont {Brunel}}, \ and\ \bibinfo
  {author} {\bibfnamefont {J.}~\bibnamefont {van Tol}},\ }\href {\doibase
  10.1063/1.2801700} {\bibfield  {journal} {\bibinfo  {journal} {Applied
  Physics Letters}\ }\textbf {\bibinfo {volume} {91}},\ \bibinfo {pages}
  {174102} (\bibinfo {year} {2007})},\ \Eprint
  {http://arxiv.org/abs/http://dx.doi.org/10.1063/1.2801700}
  {http://dx.doi.org/10.1063/1.2801700} \BibitemShut {NoStop}%
\bibitem [{\citenamefont {Takahashi}\ \emph {et~al.}(2012)\citenamefont
  {Takahashi}, \citenamefont {Brunel}, \citenamefont {Edwards}, \citenamefont
  {van Tol}, \citenamefont {Ramian}, \citenamefont {Han},\ and\ \citenamefont
  {Sherwin}}]{Takahashi_Pulsed_2012}%
  \BibitemOpen
  \bibfield  {author} {\bibinfo {author} {\bibfnamefont {S.}~\bibnamefont
  {Takahashi}}, \bibinfo {author} {\bibfnamefont {L.~C.}\ \bibnamefont
  {Brunel}}, \bibinfo {author} {\bibfnamefont {D.~T.}\ \bibnamefont {Edwards}},
  \bibinfo {author} {\bibfnamefont {J.}~\bibnamefont {van Tol}}, \bibinfo
  {author} {\bibfnamefont {G.}~\bibnamefont {Ramian}}, \bibinfo {author}
  {\bibfnamefont {S.}~\bibnamefont {Han}}, \ and\ \bibinfo {author}
  {\bibfnamefont {M.~S.}\ \bibnamefont {Sherwin}},\ }\href {\doibase
  10.1038/nature11437} {\bibfield  {journal} {\bibinfo  {journal} {Nature}\
  }\textbf {\bibinfo {volume} {489}},\ \bibinfo {pages} {409} (\bibinfo {year}
  {2012})}\BibitemShut {NoStop}%
\bibitem [{\citenamefont {Edwards}\ \emph {et~al.}(2013)\citenamefont
  {Edwards}, \citenamefont {Zhang}, \citenamefont {Glaser}, \citenamefont
  {Han},\ and\ \citenamefont {Sherwin}}]{Edwards2013}%
  \BibitemOpen
  \bibfield  {author} {\bibinfo {author} {\bibfnamefont {D.~T.}\ \bibnamefont
  {Edwards}}, \bibinfo {author} {\bibfnamefont {Y.}~\bibnamefont {Zhang}},
  \bibinfo {author} {\bibfnamefont {S.~J.}\ \bibnamefont {Glaser}}, \bibinfo
  {author} {\bibfnamefont {S.}~\bibnamefont {Han}}, \ and\ \bibinfo {author}
  {\bibfnamefont {M.~S.}\ \bibnamefont {Sherwin}},\ }\href {\doibase
  10.1039/C3CP44492A} {\bibfield  {journal} {\bibinfo  {journal} {Phys. Chem.
  Chem. Phys.}\ }\textbf {\bibinfo {volume} {15}},\ \bibinfo {pages} {5707}
  (\bibinfo {year} {2013})}\BibitemShut {NoStop}%
\bibitem [{\citenamefont {Wilson}\ \emph {et~al.}(2018)\citenamefont {Wilson},
  \citenamefont {Aronson}, \citenamefont {Clayton}, \citenamefont {Glaser},
  \citenamefont {Han},\ and\ \citenamefont {Sherwin}}]{Wilson_MultiStep_2018}%
  \BibitemOpen
  \bibfield  {author} {\bibinfo {author} {\bibfnamefont {C.~B.}\ \bibnamefont
  {Wilson}}, \bibinfo {author} {\bibfnamefont {S.}~\bibnamefont {Aronson}},
  \bibinfo {author} {\bibfnamefont {J.~A.}\ \bibnamefont {Clayton}}, \bibinfo
  {author} {\bibfnamefont {S.~J.}\ \bibnamefont {Glaser}}, \bibinfo {author}
  {\bibfnamefont {S.}~\bibnamefont {Han}}, \ and\ \bibinfo {author}
  {\bibfnamefont {M.~S.}\ \bibnamefont {Sherwin}},\ }\href {\doibase
  10.1039/C8CP01876F} {\bibfield  {journal} {\bibinfo  {journal} {Phys. Chem.
  Chem. Phys.}\ }\textbf {\bibinfo {volume} {20}},\ \bibinfo {pages} {18097}
  (\bibinfo {year} {2018})}\BibitemShut {NoStop}%
\bibitem [{\citenamefont {Rabi}(1937)}]{Rabi_Space_1937}%
  \BibitemOpen
  \bibfield  {author} {\bibinfo {author} {\bibfnamefont {I.~I.}\ \bibnamefont
  {Rabi}},\ }\href {\doibase 10.1103/PhysRev.51.652} {\bibfield  {journal}
  {\bibinfo  {journal} {Phys. Rev.}\ }\textbf {\bibinfo {volume} {51}},\
  \bibinfo {pages} {652} (\bibinfo {year} {1937})}\BibitemShut {NoStop}%
\bibitem [{\citenamefont {Kittel}(1948)}]{Kittel_Ferro_1948}%
  \BibitemOpen
  \bibfield  {author} {\bibinfo {author} {\bibfnamefont {C.}~\bibnamefont
  {Kittel}},\ }\href {\doibase 10.1103/PhysRev.73.155} {\bibfield  {journal}
  {\bibinfo  {journal} {Phys. Rev.}\ }\textbf {\bibinfo {volume} {73}},\
  \bibinfo {pages} {155} (\bibinfo {year} {1948})}\BibitemShut {NoStop}%
\bibitem [{\citenamefont {Zangwill}(2013)}]{Zangwill_Electrodynamics_2013}%
  \BibitemOpen
  \bibfield  {author} {\bibinfo {author} {\bibfnamefont {A.}~\bibnamefont
  {Zangwill}},\ }\href@noop {} {\emph {\bibinfo {title} {Modern
  Electrodynamics}}},\ \bibinfo {edition} {1st}\ ed.\ (\bibinfo  {publisher}
  {Cambridge University Press},\ \bibinfo {year} {2013})\BibitemShut {NoStop}%
\bibitem [{\citenamefont {Bloch}(1946)}]{Bloch_Nuclear_1946}%
  \BibitemOpen
  \bibfield  {author} {\bibinfo {author} {\bibfnamefont {F.}~\bibnamefont
  {Bloch}},\ }\href {\doibase 10.1103/PhysRev.70.460} {\bibfield  {journal}
  {\bibinfo  {journal} {Physical Review}\ }\textbf {\bibinfo {volume} {70}},\
  \bibinfo {pages} {460} (\bibinfo {year} {1946})}\BibitemShut {NoStop}%
\bibitem [{SI()}]{SI}%
  \BibitemOpen
  \href@noop {} {}\bibinfo {note} {See Supplemental Material at [url will be
  submitted by publisher] for details on temperature dependence, on detuning
  dependence, on measurements of relaxation times, and on numerical
  simulations.}\BibitemShut {Stop}%
\bibitem [{\citenamefont {{Suhl}}(1956)}]{Suhl_1956}%
  \BibitemOpen
  \bibfield  {author} {\bibinfo {author} {\bibfnamefont {H.}~\bibnamefont
  {{Suhl}}},\ }\href {\doibase 10.1109/JRPROC.1956.274950} {\bibfield
  {journal} {\bibinfo  {journal} {Proceedings of the IRE}\ }\textbf {\bibinfo
  {volume} {44}},\ \bibinfo {pages} {1270} (\bibinfo {year}
  {1956})}\BibitemShut {NoStop}%
\bibitem [{\citenamefont {Levitt}(1996)}]{Levitt_Demagnetization_1996}%
  \BibitemOpen
  \bibfield  {author} {\bibinfo {author} {\bibfnamefont {M.~H.}\ \bibnamefont
  {Levitt}},\ }\href {\doibase
  10.1002/(SICI)1099-0534(1996)8:2<77::AID-CMR1>3.0.CO;2-L} {\bibfield
  {journal} {\bibinfo  {journal} {Concepts in Magnetic Resonance}\ }\textbf
  {\bibinfo {volume} {8}},\ \bibinfo {pages} {77} (\bibinfo {year}
  {1996})}\BibitemShut {NoStop}%
\bibitem [{\citenamefont {Bowtell}\ \emph {et~al.}(1990)\citenamefont
  {Bowtell}, \citenamefont {Bowley},\ and\ \citenamefont
  {Glover}}]{Bowtell_Multiple_1969}%
  \BibitemOpen
  \bibfield  {author} {\bibinfo {author} {\bibfnamefont {R.}~\bibnamefont
  {Bowtell}}, \bibinfo {author} {\bibfnamefont {R.}~\bibnamefont {Bowley}}, \
  and\ \bibinfo {author} {\bibfnamefont {P.}~\bibnamefont {Glover}},\ }\href
  {\doibase https://doi.org/10.1016/0022-2364(90)90297-M} {\bibfield  {journal}
  {\bibinfo  {journal} {Journal of Magnetic Resonance (1969)}\ }\textbf
  {\bibinfo {volume} {88}},\ \bibinfo {pages} {643 } (\bibinfo {year}
  {1990})}\BibitemShut {NoStop}%
\bibitem [{\citenamefont {Deville}\ \emph {et~al.}(1979)\citenamefont
  {Deville}, \citenamefont {Bernier},\ and\ \citenamefont
  {Delrieux}}]{Denville_NMR_1979}%
  \BibitemOpen
  \bibfield  {author} {\bibinfo {author} {\bibfnamefont {G.}~\bibnamefont
  {Deville}}, \bibinfo {author} {\bibfnamefont {M.}~\bibnamefont {Bernier}}, \
  and\ \bibinfo {author} {\bibfnamefont {J.~M.}\ \bibnamefont {Delrieux}},\
  }\href {\doibase 10.1103/PhysRevB.19.5666} {\bibfield  {journal} {\bibinfo
  {journal} {Phys. Rev. B}\ }\textbf {\bibinfo {volume} {19}},\ \bibinfo
  {pages} {5666} (\bibinfo {year} {1979})}\BibitemShut {NoStop}%
\bibitem [{\citenamefont {Bedford}\ \emph {et~al.}(1991)\citenamefont
  {Bedford}, \citenamefont {Bowtell},\ and\ \citenamefont
  {Bowley}}]{Bedford_Multiple_1991}%
  \BibitemOpen
  \bibfield  {author} {\bibinfo {author} {\bibfnamefont {A.}~\bibnamefont
  {Bedford}}, \bibinfo {author} {\bibfnamefont {R.}~\bibnamefont {Bowtell}}, \
  and\ \bibinfo {author} {\bibfnamefont {R.}~\bibnamefont {Bowley}},\
  }\href@noop {} {\bibfield  {journal} {\bibinfo  {journal} {Journal of
  Magnetic Resonance (1969)}\ }\textbf {\bibinfo {volume} {93}},\ \bibinfo
  {pages} {516 } (\bibinfo {year} {1991})}\BibitemShut {NoStop}%
\bibitem [{\citenamefont {Warren}\ \emph {et~al.}(1993)\citenamefont {Warren},
  \citenamefont {Richter}, \citenamefont {Andreotti},\ and\ \citenamefont
  {Farmer}}]{Warren_Generation_1993}%
  \BibitemOpen
  \bibfield  {author} {\bibinfo {author} {\bibfnamefont {W.}~\bibnamefont
  {Warren}}, \bibinfo {author} {\bibfnamefont {W.}~\bibnamefont {Richter}},
  \bibinfo {author} {\bibfnamefont {A.}~\bibnamefont {Andreotti}}, \ and\
  \bibinfo {author} {\bibfnamefont {B.}~\bibnamefont {Farmer}},\ }\href
  {\doibase 10.1126/science.8266096} {\bibfield  {journal} {\bibinfo  {journal}
  {Science}\ }\textbf {\bibinfo {volume} {262}},\ \bibinfo {pages} {2005}
  (\bibinfo {year} {1993})}\BibitemShut {NoStop}%
\bibitem [{\citenamefont {Richter}\ \emph {et~al.}(1995)\citenamefont
  {Richter}, \citenamefont {Lee}, \citenamefont {Warren},\ and\ \citenamefont
  {He}}]{Richter_Imaging_1995}%
  \BibitemOpen
  \bibfield  {author} {\bibinfo {author} {\bibfnamefont {W.}~\bibnamefont
  {Richter}}, \bibinfo {author} {\bibfnamefont {S.}~\bibnamefont {Lee}},
  \bibinfo {author} {\bibfnamefont {W.}~\bibnamefont {Warren}}, \ and\ \bibinfo
  {author} {\bibfnamefont {Q.}~\bibnamefont {He}},\ }\href {\doibase
  10.1126/science.7839140} {\bibfield  {journal} {\bibinfo  {journal}
  {Science}\ }\textbf {\bibinfo {volume} {267}},\ \bibinfo {pages} {654}
  (\bibinfo {year} {1995})}\BibitemShut {NoStop}%
\bibitem [{\citenamefont {Jeener}(2000)}]{Jeener_Equivalence_2000}%
  \BibitemOpen
  \bibfield  {author} {\bibinfo {author} {\bibfnamefont {J.}~\bibnamefont
  {Jeener}},\ }\href {\doibase 10.1063/1.481063} {\bibfield  {journal}
  {\bibinfo  {journal} {The Journal of Chemical Physics}\ }\textbf {\bibinfo
  {volume} {112}},\ \bibinfo {pages} {5091} (\bibinfo {year}
  {2000})}\BibitemShut {NoStop}%
\bibitem [{\citenamefont {Edzes}(1990)}]{Edzes_TheNuclear_1969}%
  \BibitemOpen
  \bibfield  {author} {\bibinfo {author} {\bibfnamefont {H.~T.}\ \bibnamefont
  {Edzes}},\ }\href {\doibase https://doi.org/10.1016/0022-2364(90)90261-7}
  {\bibfield  {journal} {\bibinfo  {journal} {Journal of Magnetic Resonance
  (1969)}\ }\textbf {\bibinfo {volume} {86}},\ \bibinfo {pages} {293 }
  (\bibinfo {year} {1990})}\BibitemShut {NoStop}%
\bibitem [{\citenamefont {Bloembergen}\ and\ \citenamefont
  {Pound}(1954)}]{Bloembergen_Radiation_1954}%
  \BibitemOpen
  \bibfield  {author} {\bibinfo {author} {\bibfnamefont {N.}~\bibnamefont
  {Bloembergen}}\ and\ \bibinfo {author} {\bibfnamefont {R.~V.}\ \bibnamefont
  {Pound}},\ }\href {\doibase 10.1103/PhysRev.95.8} {\bibfield  {journal}
  {\bibinfo  {journal} {Phys. Rev.}\ }\textbf {\bibinfo {volume} {95}},\
  \bibinfo {pages} {8} (\bibinfo {year} {1954})}\BibitemShut {NoStop}%
\bibitem [{\citenamefont {Augustine}(2002)}]{Augustine_Transient_2002}%
  \BibitemOpen
  \bibfield  {author} {\bibinfo {author} {\bibfnamefont {M.}~\bibnamefont
  {Augustine}},\ }\href {\doibase 10.1016/S0079-6565(01)00037-1} {\bibfield
  {journal} {\bibinfo  {journal} {Progress in Nuclear Magnetic Resonance
  Spectroscopy - PROG NUCL MAGN RESON SPECTROS}\ }\textbf {\bibinfo {volume}
  {40}} (\bibinfo {year} {2002}),\ 10.1016/S0079-6565(01)00037-1}\BibitemShut
  {NoStop}%
\bibitem [{\citenamefont {Benner}(1977)}]{Benner_Influence_977}%
  \BibitemOpen
  \bibfield  {author} {\bibinfo {author} {\bibfnamefont {H.}~\bibnamefont
  {Benner}},\ }\href {\doibase 10.1007/BF00882472} {\bibfield  {journal}
  {\bibinfo  {journal} {Applied physics}\ }\textbf {\bibinfo {volume} {13}},\
  \bibinfo {pages} {141} (\bibinfo {year} {1977})}\BibitemShut {NoStop}%
\bibitem [{\citenamefont {Prisner}\ \emph {et~al.}(1992)\citenamefont
  {Prisner}, \citenamefont {Un},\ and\ \citenamefont
  {Griffin}}]{Prisner_Pulsed_1992}%
  \BibitemOpen
  \bibfield  {author} {\bibinfo {author} {\bibfnamefont {T.}~\bibnamefont
  {Prisner}}, \bibinfo {author} {\bibfnamefont {S.}~\bibnamefont {Un}}, \ and\
  \bibinfo {author} {\bibfnamefont {R.}~\bibnamefont {Griffin}},\ }\href
  {\doibase 10.1002/ijch.199200042} {\bibfield  {journal} {\bibinfo  {journal}
  {Israel Journal of Chemistry}\ }\textbf {\bibinfo {volume} {32}},\ \bibinfo
  {pages} {357} (\bibinfo {year} {1992})},\ \Eprint
  {http://arxiv.org/abs/https://onlinelibrary.wiley.com/doi/pdf/10.1002/ijch.199200042}
  {https://onlinelibrary.wiley.com/doi/pdf/10.1002/ijch.199200042} \BibitemShut
  {NoStop}%
\bibitem [{\citenamefont {Kitagawa}\ and\ \citenamefont
  {Ueda}(1993)}]{Kitagawa_Squeezed_1993}%
  \BibitemOpen
  \bibfield  {author} {\bibinfo {author} {\bibfnamefont {M.}~\bibnamefont
  {Kitagawa}}\ and\ \bibinfo {author} {\bibfnamefont {M.}~\bibnamefont
  {Ueda}},\ }\href {\doibase 10.1103/PhysRevA.47.5138} {\bibfield  {journal}
  {\bibinfo  {journal} {Phys. Rev. A}\ }\textbf {\bibinfo {volume} {47}},\
  \bibinfo {pages} {5138} (\bibinfo {year} {1993})}\BibitemShut {NoStop}%
\bibitem [{\citenamefont {Bennett}\ \emph {et~al.}(2013)\citenamefont
  {Bennett}, \citenamefont {Yao}, \citenamefont {Otterbach}, \citenamefont
  {Zoller}, \citenamefont {Rabl},\ and\ \citenamefont
  {Lukin}}]{Bennet_Phonon_2013}%
  \BibitemOpen
  \bibfield  {author} {\bibinfo {author} {\bibfnamefont {S.~D.}\ \bibnamefont
  {Bennett}}, \bibinfo {author} {\bibfnamefont {N.~Y.}\ \bibnamefont {Yao}},
  \bibinfo {author} {\bibfnamefont {J.}~\bibnamefont {Otterbach}}, \bibinfo
  {author} {\bibfnamefont {P.}~\bibnamefont {Zoller}}, \bibinfo {author}
  {\bibfnamefont {P.}~\bibnamefont {Rabl}}, \ and\ \bibinfo {author}
  {\bibfnamefont {M.~D.}\ \bibnamefont {Lukin}},\ }\href {\doibase
  10.1103/PhysRevLett.110.156402} {\bibfield  {journal} {\bibinfo  {journal}
  {Phys. Rev. Lett.}\ }\textbf {\bibinfo {volume} {110}},\ \bibinfo {pages}
  {156402} (\bibinfo {year} {2013})}\BibitemShut {NoStop}%
\bibitem [{\citenamefont {Ma}\ \emph {et~al.}(2016)\citenamefont {Ma},
  \citenamefont {Zhang}, \citenamefont {Song},\ and\ \citenamefont
  {Wu}}]{Ma_Bistability_2016}%
  \BibitemOpen
  \bibfield  {author} {\bibinfo {author} {\bibfnamefont {Y.-H.}\ \bibnamefont
  {Ma}}, \bibinfo {author} {\bibfnamefont {X.-F.}\ \bibnamefont {Zhang}},
  \bibinfo {author} {\bibfnamefont {J.}~\bibnamefont {Song}}, \ and\ \bibinfo
  {author} {\bibfnamefont {E.}~\bibnamefont {Wu}},\ }\href {\doibase
  https://doi.org/10.1016/j.aop.2016.03.001} {\bibfield  {journal} {\bibinfo
  {journal} {Annals of Physics}\ }\textbf {\bibinfo {volume} {369}},\ \bibinfo
  {pages} {36 } (\bibinfo {year} {2016})}\BibitemShut {NoStop}%
\bibitem [{\citenamefont {Trifunovic}\ \emph {et~al.}(2013)\citenamefont
  {Trifunovic}, \citenamefont {Pedrocchi},\ and\ \citenamefont
  {Loss}}]{Trifunovic_2013}%
  \BibitemOpen
  \bibfield  {author} {\bibinfo {author} {\bibfnamefont {L.}~\bibnamefont
  {Trifunovic}}, \bibinfo {author} {\bibfnamefont {F.~L.}\ \bibnamefont
  {Pedrocchi}}, \ and\ \bibinfo {author} {\bibfnamefont {D.}~\bibnamefont
  {Loss}},\ }\href {\doibase 10.1103/PhysRevX.3.041023} {\bibfield  {journal}
  {\bibinfo  {journal} {Phys. Rev. X}\ }\textbf {\bibinfo {volume} {3}},\
  \bibinfo {pages} {041023} (\bibinfo {year} {2013})}\BibitemShut {NoStop}%
\end{thebibliography}%


\begin{thebibliography}{8}%
\makeatletter
\providecommand \@ifxundefined [1]{%
 \@ifx{#1\undefined}
}%
\providecommand \@ifnum [1]{%
 \ifnum #1\expandafter \@firstoftwo
 \else \expandafter \@secondoftwo
 \fi
}%
\providecommand \@ifx [1]{%
 \ifx #1\expandafter \@firstoftwo
 \else \expandafter \@secondoftwo
 \fi
}%
\providecommand \natexlab [1]{#1}%
\providecommand \enquote  [1]{``#1''}%
\providecommand \bibnamefont  [1]{#1}%
\providecommand \bibfnamefont [1]{#1}%
\providecommand \citenamefont [1]{#1}%
\providecommand \href@noop [0]{\@secondoftwo}%
\providecommand \href [0]{\begingroup \@sanitize@url \@href}%
\providecommand \@href[1]{\@@startlink{#1}\@@href}%
\providecommand \@@href[1]{\endgroup#1\@@endlink}%
\providecommand \@sanitize@url [0]{\catcode `\\12\catcode `\$12\catcode
  `\&12\catcode `\#12\catcode `\^12\catcode `\_12\catcode `\%12\relax}%
\providecommand \@@startlink[1]{}%
\providecommand \@@endlink[0]{}%
\providecommand \url  [0]{\begingroup\@sanitize@url \@url }%
\providecommand \@url [1]{\endgroup\@href {#1}{\urlprefix }}%
\providecommand \urlprefix  [0]{URL }%
\providecommand \Eprint [0]{\href }%
\providecommand \doibase [0]{http://dx.doi.org/}%
\providecommand \selectlanguage [0]{\@gobble}%
\providecommand \bibinfo  [0]{\@secondoftwo}%
\providecommand \bibfield  [0]{\@secondoftwo}%
\providecommand \translation [1]{[#1]}%
\providecommand \BibitemOpen [0]{}%
\providecommand \bibitemStop [0]{}%
\providecommand \bibitemNoStop [0]{.\EOS\space}%
\providecommand \EOS [0]{\spacefactor3000\relax}%
\providecommand \BibitemShut  [1]{\csname bibitem#1\endcsname}%
\let\auto@bib@innerbib\@empty
\bibitem [{\citenamefont {Wilson}\ \emph {et~al.}(2018)\citenamefont {Wilson},
  \citenamefont {Aronson}, \citenamefont {Clayton}, \citenamefont {Glaser},
  \citenamefont {Han},\ and\ \citenamefont {Sherwin}}]{Wilson_MultiStep_2018}%
  \BibitemOpen
  \bibfield  {author} {\bibinfo {author} {\bibfnamefont {C.~B.}\ \bibnamefont
  {Wilson}}, \bibinfo {author} {\bibfnamefont {S.}~\bibnamefont {Aronson}},
  \bibinfo {author} {\bibfnamefont {J.~A.}\ \bibnamefont {Clayton}}, \bibinfo
  {author} {\bibfnamefont {S.~J.}\ \bibnamefont {Glaser}}, \bibinfo {author}
  {\bibfnamefont {S.}~\bibnamefont {Han}}, \ and\ \bibinfo {author}
  {\bibfnamefont {M.~S.}\ \bibnamefont {Sherwin}},\ }\href {\doibase
  10.1039/C8CP01876F} {\bibfield  {journal} {\bibinfo  {journal} {Phys. Chem.
  Chem. Phys.}\ }\textbf {\bibinfo {volume} {20}},\ \bibinfo {pages} {18097}
  (\bibinfo {year} {2018})}\BibitemShut {NoStop}%
\bibitem [{\citenamefont {Klauder}\ and\ \citenamefont
  {Anderson}(1962)}]{Klauder1962}%
  \BibitemOpen
  \bibfield  {author} {\bibinfo {author} {\bibfnamefont {J.~R.}\ \bibnamefont
  {Klauder}}\ and\ \bibinfo {author} {\bibfnamefont {P.~W.}\ \bibnamefont
  {Anderson}},\ }\href {\doibase 10.1103/PhysRev.125.912} {\bibfield  {journal}
  {\bibinfo  {journal} {Phys. Rev.}\ }\textbf {\bibinfo {volume} {125}},\
  \bibinfo {pages} {912} (\bibinfo {year} {1962})}\BibitemShut {NoStop}%
\bibitem [{\citenamefont {Salikhov}\ \emph {et~al.}(1981)\citenamefont
  {Salikhov}, \citenamefont {Dzuba},\ and\ \citenamefont
  {Raitsimring}}]{Salikhov1981}%
  \BibitemOpen
  \bibfield  {author} {\bibinfo {author} {\bibfnamefont {K.}~\bibnamefont
  {Salikhov}}, \bibinfo {author} {\bibfnamefont {S.}~\bibnamefont {Dzuba}}, \
  and\ \bibinfo {author} {\bibfnamefont {A.}~\bibnamefont {Raitsimring}},\
  }\href {\doibase http://dx.doi.org/10.1016/0022-2364(81)90216-X} {\bibfield
  {journal} {\bibinfo  {journal} {Journal of Magnetic Resonance (1969)}\
  }\textbf {\bibinfo {volume} {42}},\ \bibinfo {pages} {255 } (\bibinfo {year}
  {1981})}\BibitemShut {NoStop}%
\bibitem [{\citenamefont {Schweiger}\ and\ \citenamefont
  {Jeschke}(2001)}]{Jeschke_Electron_2001}%
  \BibitemOpen
  \bibfield  {author} {\bibinfo {author} {\bibfnamefont {A.}~\bibnamefont
  {Schweiger}}\ and\ \bibinfo {author} {\bibfnamefont {G.}~\bibnamefont
  {Jeschke}},\ }\href@noop {} {\emph {\bibinfo {title} {Principles of pulse
  electron paramagnetic resonance}}}\ (\bibinfo  {publisher} {Oxford University
  Press},\ \bibinfo {year} {2001})\BibitemShut {NoStop}%
\bibitem [{\citenamefont {Duffy}\ \emph {et~al.}(1972)\citenamefont {Duffy},
  \citenamefont {Dubach}, \citenamefont {Pianetta}, \citenamefont {Deck},
  \citenamefont {Strandburg},\ and\ \citenamefont
  {Miedema}}]{Duffy_Antiferromagnetic_1972}%
  \BibitemOpen
  \bibfield  {author} {\bibinfo {author} {\bibfnamefont {W.}~\bibnamefont
  {Duffy}}, \bibinfo {author} {\bibfnamefont {J.~F.}\ \bibnamefont {Dubach}},
  \bibinfo {author} {\bibfnamefont {P.~A.}\ \bibnamefont {Pianetta}}, \bibinfo
  {author} {\bibfnamefont {J.~F.}\ \bibnamefont {Deck}}, \bibinfo {author}
  {\bibfnamefont {D.~L.}\ \bibnamefont {Strandburg}}, \ and\ \bibinfo {author}
  {\bibfnamefont {A.~R.}\ \bibnamefont {Miedema}},\ }\href {\doibase
  doi:http://dx.doi.org/10.1063/1.1677580} {\bibfield  {journal} {\bibinfo
  {journal} {The Journal of Chemical Physics}\ }\textbf {\bibinfo {volume}
  {56}},\ \bibinfo {pages} {2555} (\bibinfo {year} {1972})}\BibitemShut
  {NoStop}%
\bibitem [{\citenamefont {Bloembergen}\ and\ \citenamefont
  {Pound}(1954)}]{Bloembergen_Radiation_1954}%
  \BibitemOpen
  \bibfield  {author} {\bibinfo {author} {\bibfnamefont {N.}~\bibnamefont
  {Bloembergen}}\ and\ \bibinfo {author} {\bibfnamefont {R.~V.}\ \bibnamefont
  {Pound}},\ }\href {\doibase 10.1103/PhysRev.95.8} {\bibfield  {journal}
  {\bibinfo  {journal} {Phys. Rev.}\ }\textbf {\bibinfo {volume} {95}},\
  \bibinfo {pages} {8} (\bibinfo {year} {1954})}\BibitemShut {NoStop}%
\bibitem [{\citenamefont {BLOOM}(1957)}]{Bloom_Effects_1957}%
  \BibitemOpen
  \bibfield  {author} {\bibinfo {author} {\bibfnamefont {S.}~\bibnamefont
  {BLOOM}},\ }\href {\doibase 10.1063/1.1722859} {\bibfield  {journal}
  {\bibinfo  {journal} {JOURNAL OF APPLIED PHYSICS}\ }\textbf {\bibinfo
  {volume} {28}},\ \bibinfo {pages} {800} (\bibinfo {year} {1957})}\BibitemShut
  {NoStop}%
\bibitem [{\citenamefont {Augustine}(2002)}]{Augustine_Transient_2002}%
  \BibitemOpen
  \bibfield  {author} {\bibinfo {author} {\bibfnamefont {M.}~\bibnamefont
  {Augustine}},\ }\href {\doibase 10.1016/S0079-6565(01)00037-1} {\bibfield
  {journal} {\bibinfo  {journal} {Progress in Nuclear Magnetic Resonance
  Spectroscopy - PROG NUCL MAGN RESON SPECTROS}\ }\textbf {\bibinfo {volume}
  {40}} (\bibinfo {year} {2002}),\ 10.1016/S0079-6565(01)00037-1}\BibitemShut
  {NoStop}%
\end{thebibliography}%

\end{document}



\title{Supplemental Material for ``Dressed Rabi oscillation in a crystalline organic radical''}
\author{C. Blake Wilson}
\affiliation{Department of Physics, University of California, Santa Barbara, Santa Barbara, California, USA}
\affiliation{Institute for Terahertz Science and Technology, University of California, Santa Barbara, Santa Barbara, California, USA.}

\author{Devin T. Edwards}
\affiliation{Department of Physics, University of California, Santa Barbara, Santa Barbara, California, USA}
\affiliation{Institute for Terahertz Science and Technology, University of California, Santa Barbara, Santa Barbara, California, USA.}

\author{Jessica A. Clayton}
\affiliation{Department of Physics, University of California, Santa Barbara, Santa Barbara, California, USA}
\affiliation{Institute for Terahertz Science and Technology, University of California, Santa Barbara, Santa Barbara, California, USA.}

\author{Songi Han}
\affiliation{Institute for Terahertz Science and Technology, University of California, Santa Barbara, Santa Barbara, California, USA.}
\affiliation{Department of Chemistry and Biochemistry, University of California, Santa Barbara, Santa Barbara, California, USA}

\author{Mark S. Sherwin}
\affiliation{Department of Physics, University of California, Santa Barbara, Santa Barbara, California, USA}
\affiliation{Institute for Terahertz Science and Technology, University of California, Santa Barbara, Santa Barbara, California, USA.}
%

\maketitle


\section{Temperature dependence data}

\begin{figure}[ht]
    \centering
    \includegraphics{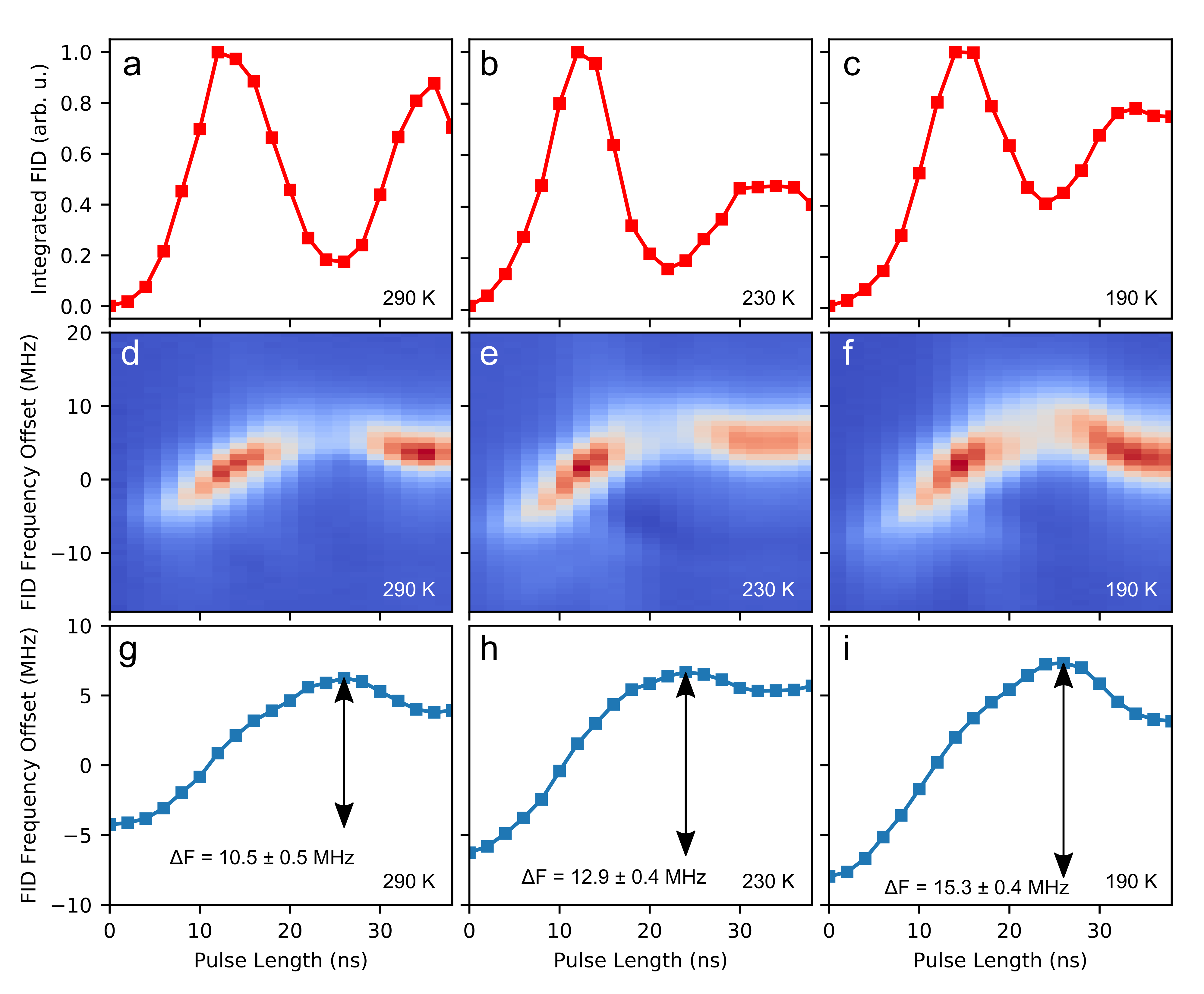}
    \caption{
    Top: Integrated FT-EPR amplitude as a function of pulse length, demonstrating Rabi oscillations at 290 K \textbf{a}, 230 K \textbf{b}, and 190 K \textbf{c}. \textbf{d}-\textbf{f}: Contour plots showing the FT-EPR absorption lineshape as a function of pulse length, showing the shift in the EPR condition. \textbf{g}-\textbf{i}: Mean FID frequency as a function of pulse length. The maximum frequency shift $\Delta$F is recorded for each temperature.
    }
    \label{fig:BigTemperatureFigure}
\end{figure}

Electron spin nutation experiments were performed on BDPA-Bz at 290 K, 230 K, and 190 K, in order to study the temperature dependence of the induced frequency shift. At all three temperatures, Rabi oscillations were observed which were accompanied by a tip-angle dependent frequency shift. Free induction decays (FIDs) were generated on a grain of BDPA-Bz and studied as a function of pulse length. Experiments were performed with a Rabi frequency $\gamma B_1/2\pi = 20$ MHz. Figure \ref{fig:BigTemperatureFigure} shows the results of temperature-dependent nutation experiments, together with the measured maximum frequency shifts.

\section{On- and off-resonance experiments}

\begin{figure}[ht]
    \centering
    \includegraphics{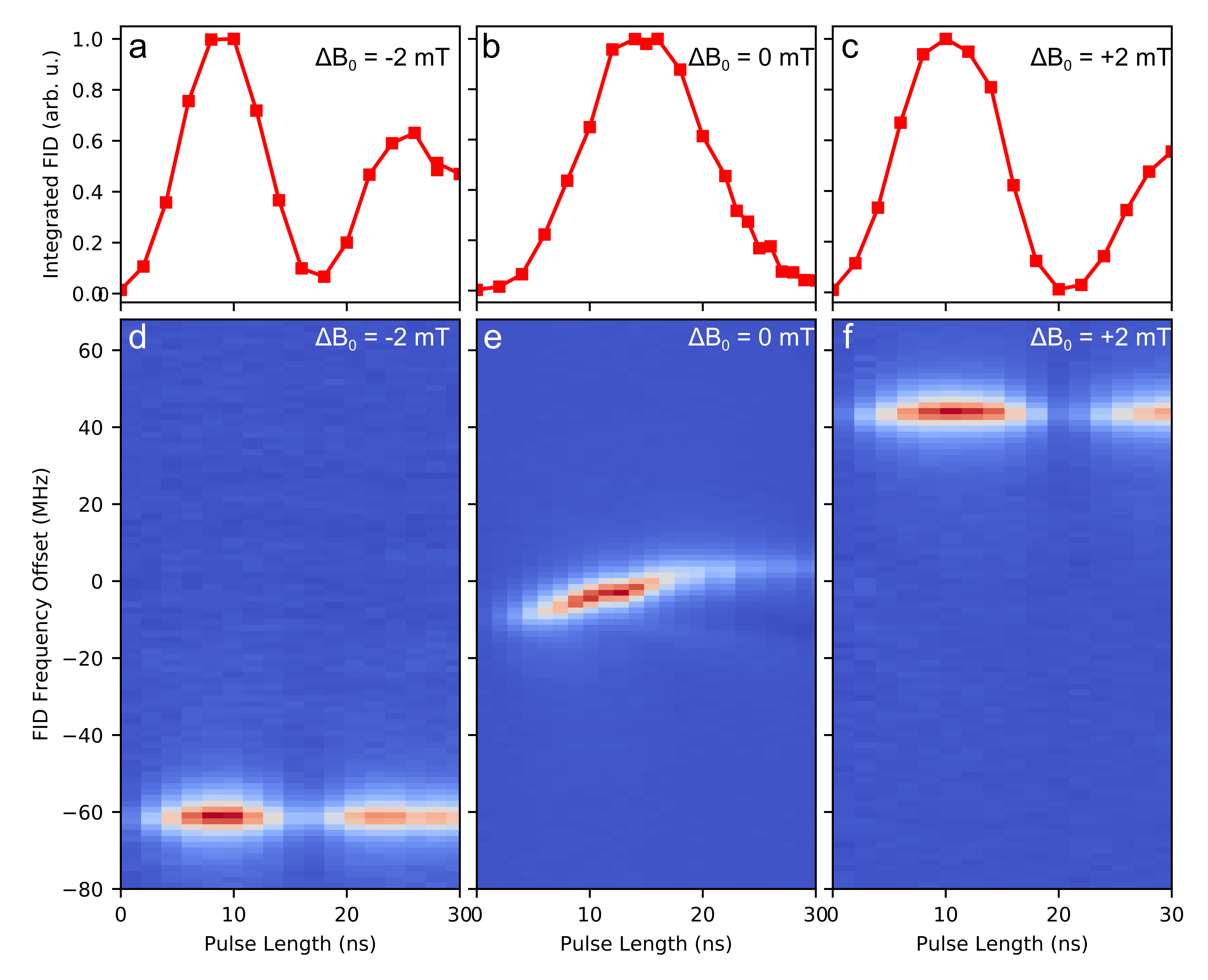}
    \caption{
    Integrated FT-EPR amplitude as a function of pulse length with the magnetic field $B_0$ set 2 mT below the Larmor condition \textbf{a}, at the Larmor condition \textbf{b}, and 2 mT above the Larmor condition \textbf{c}. \textbf{d}-\textbf{f} show the corresponding contour plots of the FT-EPR amplitude as a function of pulse length. When $B_0$ is moved away from the Larmor condition, the spin system evolves in the small tip-angle regime and no tip-angle dependent frequency shift is observed, while when $\Delta B_0 = 0$, the spin precession frequency shifts as the magnetization is inverted.
    }
    \label{fig:BigResonanceFigure}
\end{figure}

Electron spin nutation experiments were performed on BDPA-Bz at 290 K while varying the offset between electron spin resonance condition and the driving field from the FEL. Offset frequency was varied by changing the magnetic field $B_0$ at constant FEL frequency. Figure \ref{fig:BigResonanceFigure} shows the results of experiments performed with the magnetic field $B_0$ moved 2 mT above the resonance condition or 2 mT below the resonance condition. The integrated FT-EPR intensity shows a dependence on the pulse length characteristic of off-resonance nutation experiments, where the spin system nutates in the small tip-angle regime at an effective Rabi frequency $\Omega = \sqrt{(\gamma B_1)^2 + \delta^2}$, where $\delta = \gamma\Delta B_0 - \gamma\mu_0\theta_d M_0$ is the detuning from the Larmor condition and where the demagnetization field has been taken into account. When the field is detuned away from the Larmor condition by $\pm2$ mT and the magnetization remains in the small tip-angle regime, the $z$-component of the magnetization changes only slightly and the magnetization never reaches the transverse plane. Therefore, the precession frequency $\omega(M_z) = \mu_0 \gamma(H_0 - \theta_d M_z)$ remains unchanged during the experiment, leading to no anomalous frequency shift. When the field $B_0$ is set so that the FEL driving field matches the Larmor condition, the magnetization leaves the small tip-angle regime, and a demagnetization field-induced frequency shift is observed. Experiments were performed with an on-resonance Rabi frequency $\gamma B_1/2\pi = 17 MHz$.

\section{$T_1$ and $T_2$ measurements}

\begin{figure}[ht]
    \centering
    \includegraphics{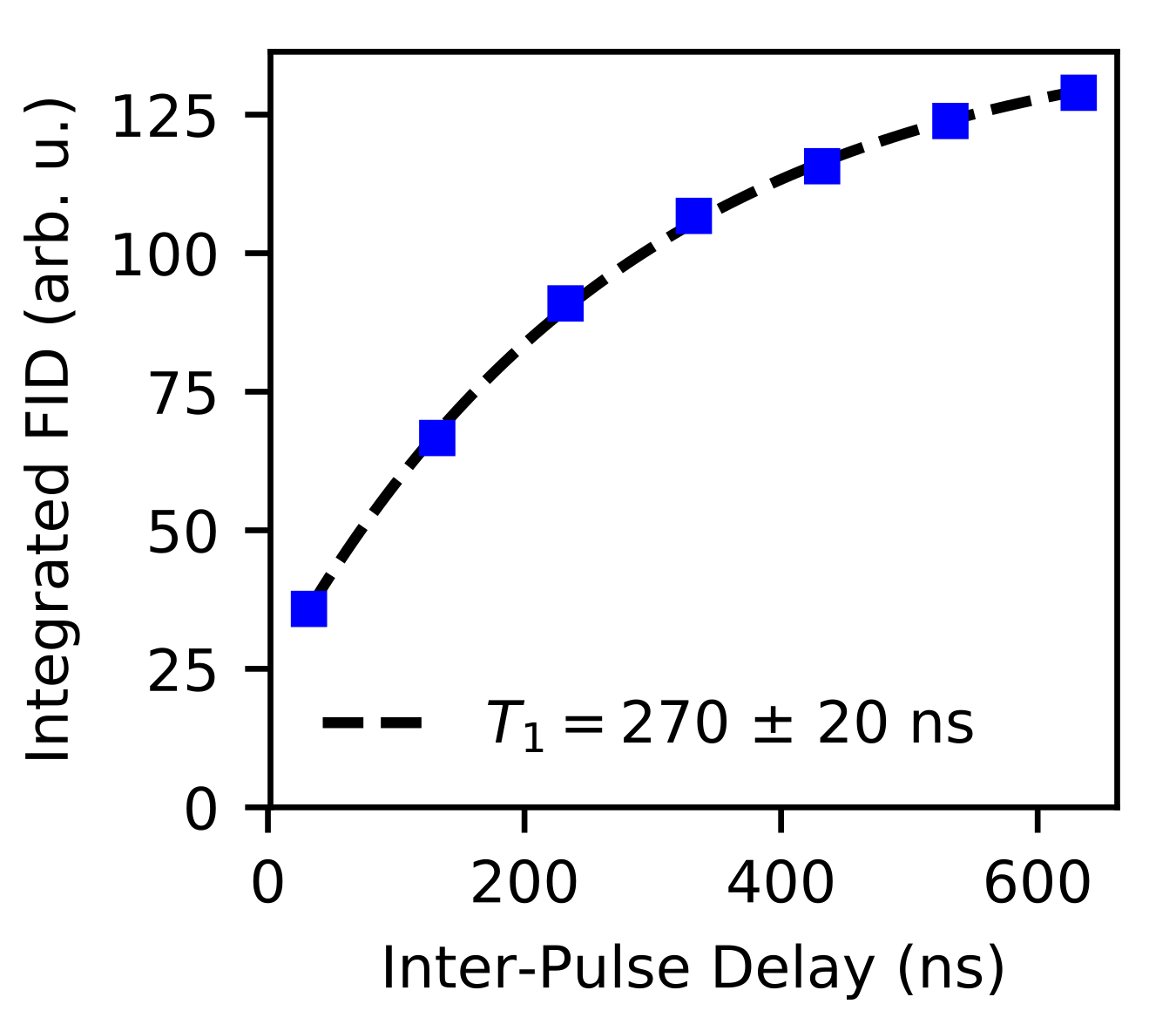}
    \caption{
    Integrated FT-EPR amplitude as a function of inter-pulse delay $T$ between the saturation pulse and the read-put pulse, the latter of which generates a FID. Dashed line shows a functional fit of the form $y(T) = C - A\times e^{-T/T_1}$, from which $T_1$ was extracted.
    }
    \label{fig:T1Figure}
\end{figure}

The phenomenological longitudinal relaxation time $T_1$ was measured using a FID-detected saturation-recovery experiment. A $\sim1.5$ $\mu$s long, 450 W pulse sliced from the long FEL pulse with an on-resonance Rabi frequency $\gamma B_1/2\pi = 8$ MHz was used to saturate the BDPA-Bz resonance. After an inter-pulse delay $T$, a 10 ns long, 5.5 kW pulse sliced from cavity dump region of the FEL pulse with an on-resonance Rabi frequency $\gamma B_1/2\pi = 25$ MHz was used to generate a FID. Four-step phase cycling was used \cite{Wilson_MultiStep_2018}. The FT-EPR lineshape was extracted from the FID. Figure \ref{fig:T1Figure} shows the integrated FT-EPR amplitude as a function of inter-pulse delay $T$, at 290 K. $T_1$ was extracted by fitting the FT-EPR amplitude $y(T)$ to a function of the form $y(T) = C - A\times e^{-T/T_1}$, where $C = y_{\infty}$ is the integrated FT-EPR absorption in the absence of an inversion pulse.

\begin{figure}[ht]
    \centering
    \includegraphics{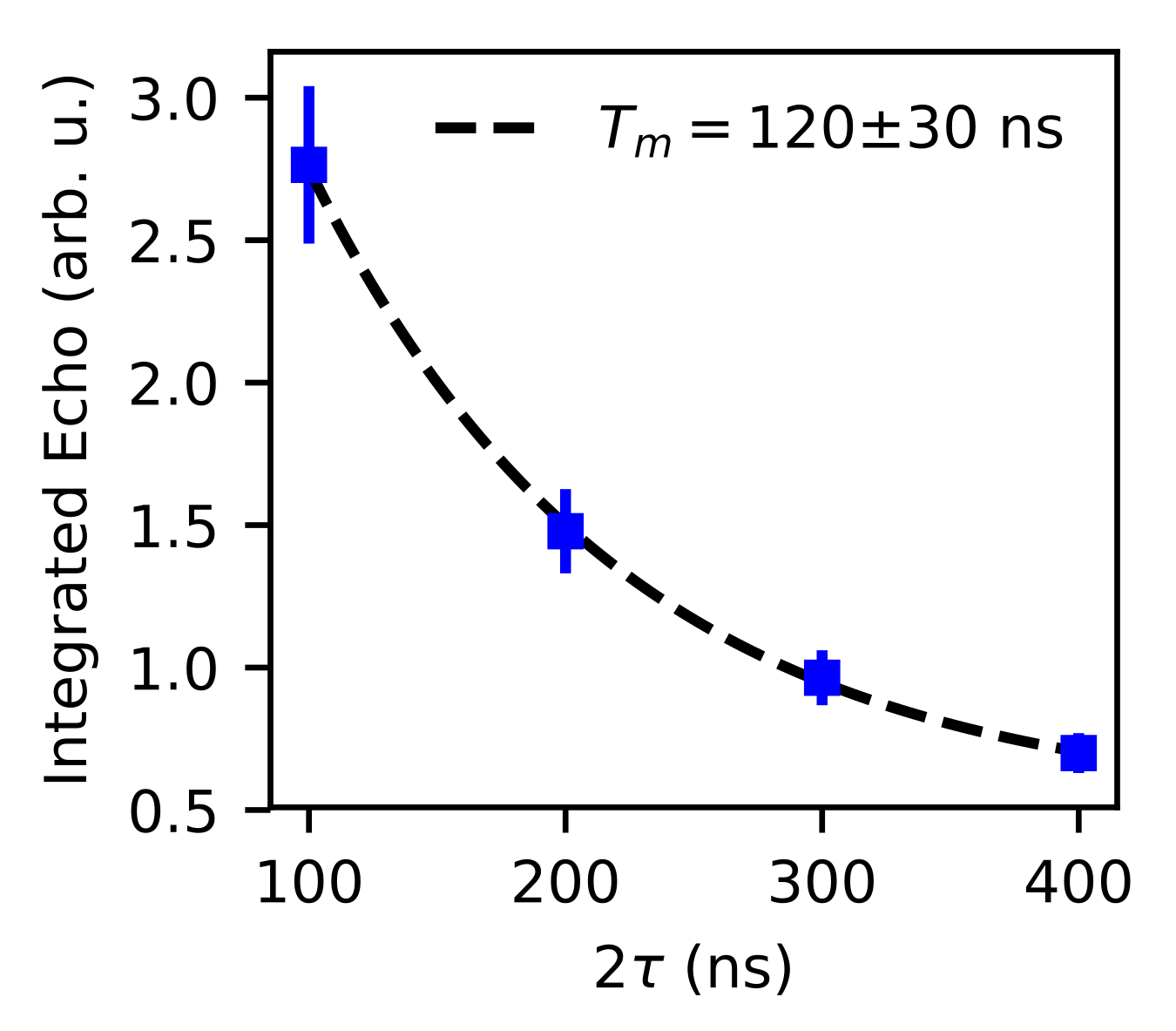}
    \caption{
    Integrated electron spin echo as a function of twice the delay $\tau$ between the first pulse and the refocusing pulse. The electron spin echo decay was fit to a function of the form $E(2\tau) = A\times e^{-2\tau/T_2} + C$, from which $T_2$ was extracted.
    }
    \label{fig:T2Figure}
\end{figure}

The transverse, spin-spin relaxation time $T_2$ was estimated using an electron spin echo decay pulse sequence of the form $P1 - \tau - P2 - \tau - echo$, where $P1$ and $P2$ were pulses of length 11.5 ns and 14 ns, respectively, $\tau$ was a delay of variable length, and $echo$ was the generated electron spin-echo. The first pulse $P1$ was sliced from the long, 450 W FEL pulse, which had an on-resonance Rabi frequency $\gamma B_1/2\pi = 8$ MHz and was therefore a small tip-angle pulse. After a variable delay $\tau$, the second pulse $P2$ was applied to refocus the magnetization. The power in the refocusing pulse $P2$ was attenuated to 55 W, for an on-resonance Rabi frequency of 2.5 MHz, in order to minimize the effects of instantaneous spectral diffusion \cite{Klauder1962, Salikhov1981, Jeschke_Electron_2001, Wilson_MultiStep_2018} and the demagnetizing field shift. Four-step phase cycling was used \cite{Wilson_MultiStep_2018}. Figure \ref{fig:T2Figure} shows the integrated echo signal $E(2\tau)$ as a function of twice the inter-pulse delay $\tau$, at 290 K. $T_2$ was extracted by fitting the echo decay to a function of the form $E(2\tau) = A\times e^{2\tau/T_2} + C$.

\section{Frequency Shift Data}

\begin{figure}[ht]
    \centering
    \includegraphics{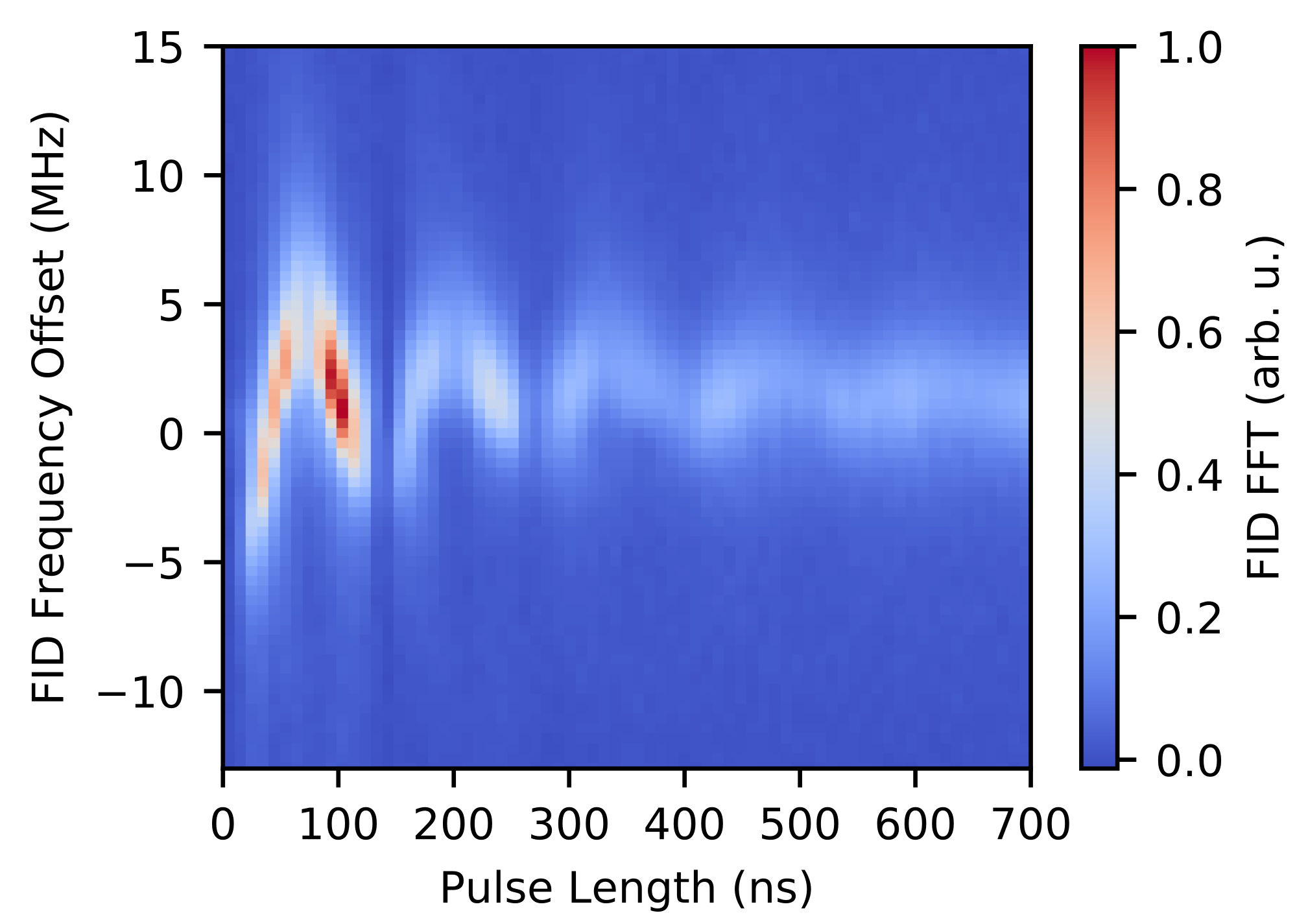}
    \caption{
    Contour plot of the FT-EPR absorption lineshape as a function of pulse length for BDPA-Bz accompanying the mean frequency shift data presented in Figure 3.
    }
    \label{fig:Long_Pulse_FID}
\end{figure}

The mean FID precision frequency was determined by fitting a complex Lorentzian to the Fourier transform of the FID. Figure \ref{fig:Long_Pulse_FID} shows a contour plot of the FT-EPR absorption lineshape as a function of pulse length, accompanying the mean frequecny data presented in Figure 3 of the manuscript.

\section{Simulation Details}

Numerical integration of the nonlinear Bloch equations was performed in Python. Inputs to the simulation included the magnetization density, the sample geometry, the phenomenological spin-lattice and spin-spin relaxation times $T_1$ and $T_2$, the detuning between the FEL frequency and the resonance frequency of the spin system, and the on-resonance Rabi frequency $\omega_1$.

The sample magnetization density was calculated to be 270 A/m at room temperature \cite{Duffy_Antiferromagnetic_1972}. BDPA-Bz grains were examined under an optical microscope, and screened for size and shape. A grain roughly 450 $\mu$m across and 65 $\mu$m thick was chosen, whose shape conformed roughly to that of an oblate spheroid with an aspect ratio of 7:1. $T_1$ and $T_2$ measurements, shown in Figures \ref{fig:T1Figure} and \ref{fig:T2Figure}, respectively, set constraints on the relaxation times. The detuning between the FEL frequency and the resonance frequency of the spin system was controlled by adjusting the magnetic field $B_0$ experienced by the spin system, for a given FEL frequency.

In the linear response regime, the on-resonance Rabi frequency can be easily measured by measuring Rabi oscillations. However, when the resonance frequency shifts with tip angle, the situation is not so straightforward, and must be found by comparing the results of a Rabi oscillation experiment to simulated magnetization trajectories. As a starting point, the on-resonance Rabi frequency $\omega_1$ was estimated from the Bloch equations to be $\omega_1 = \pi/\tau_{270-90}$, where $\tau_{270-90}$ is the time between successive maxima on a plot of integrated FT-EPR amplitude vs pulse length.

Confidence intervals were generated through a Monte Carlo procedure, taking into account the uncertainties in $T_1$, $T_2$, the detuning, and the estimated uncertainty in $\omega_1$.

\section{Energy Lost to Magnetic Dipole Radiation}

A time-varying magnetic dipole moment $\bm{\mu}$ radiates magnetic dipole radiation with total power $P$, given by
\begin{equation}
    P = \frac{\mu_0}{6\pi c^3}|\ddot{\bm{\mu}}|^2
\end{equation}
The total magnetic dipole moment $\mathbf{\mu}$ of a sample of volume $V$ with uniform magnetization $\mathbf{M}$ is given by
\begin{equation}
    \bm{\mu} = \mathbf{M}V
\end{equation}
so that the total power radiated by the sample is given by
\begin{equation}
    P = \frac{\mu_0}{6\pi c^3}V^2|\ddot{\mathbf{M}}|^2
    \label{eq:PowerFromMagneticDipoleRadiation}
\end{equation}

Ignoring $T_1$ and $T_2$ relaxation and treating the magnetization as a classical vector with fixed magnitude $M_0$, the solution to the Bloch equations gives the magnetization as $\mathbf{M} = M_0\cos\theta\hat{z} + M_0\sin\theta e^{i\omega_L t}(\hat{x} + i\hat{y})$ where $\omega_L$ is the Larmor frequency and $\theta$ is the polar angle between the magnetization and the external magnetic field $\mathbf{B_0} = B_0\hat{z}$. In the limit where Larmor precession is much faster the rate of change of the polar angle $\theta$ so that $|\omega_L| \gg |\dot{\theta}|, |\omega_L|^2 \gg |\ddot{\theta}|$,
\begin{equation}
    P = \frac{\mu_0}{6\pi c^3}V^2 M_0^2\omega_L^4\sin^2\theta
    \label{eq:PowerRadiated}
\end{equation}

The total energy $U = -\bm{\mu}\cdot\mathbf{B}$ of the spin system is given by 
\begin{equation}
    U = -M_0VB_0\cos\theta
    \label{eq:total_energy}
\end{equation}
If the spin system loses energy only through magnetic dipole radiation, so that $dU/dt = - P$,
Equations \ref{eq:PowerRadiated} and \ref{eq:total_energy} give
\begin{equation}
    \dot{\theta} = 
    -\frac{\mu_0}{6\pi c^3}\frac{V M_0}{B_0}\omega_L^4\sin\theta
\end{equation}
which has the solution 
\begin{equation}
    \theta = 2\arctan\left[\exp \left(-\frac{t - t_0}{\tau_{rad}}\right) \right]
\end{equation}
where
\begin{subequations}
\begin{align}
    \tau_{rad} &= \left( \frac{\mu_0}{6\pi c^3}\frac{V M_0}{B_0}\omega_L^4 \right)^{-1} \\
    t_0 &= \tau_{rad} \ln \left[\tan\left( \frac{\theta_0}{2} \right) \right]
\end{align}
\end{subequations}
where $\theta_0$ is the initial tip-angle. 

For an initial tip-angle $\theta = \pi/2$, the transverse component of $\mathbf{M}$, given by $\mathbf{M}_{\perp} = M_0\sin\theta e^{i\omega_L t}(\hat{x} + i\hat{y})$, decays in magnitude according to
\begin{equation}
    |\mathbf{M_{\perp}}| = M_0 \sech \left( \frac{t}{\tau_{rad}} \right)
\end{equation}

\section{Radiation Damping}

Radiation damping \cite{Bloembergen_Radiation_1954} is a back-reaction which occurs when a spin system reacts to its own radiated field \cite{Bloom_Effects_1957}. The precessing transverse magnetization produces a reaction field $\mathbf{H}_r$ which is $\pi/2$ out of phase with the magnetization, with components \cite{Augustine_Transient_2002}
\begin{subequations}
\begin{align}
    H_{r,x} &= 2\pi Q \eta M_y \\
    H_{r,y} &= -2\pi Q \eta M_x \\
    H_{r,z} &= 0
\end{align}
\end{subequations}
where $Q$ is the quality factor of the microwave circuit and $\eta$ is the filling factor \cite{Bloembergen_Radiation_1954}. 

The effect of the reaction field $\mathbf{H}_r$ is to modify the Bloch equations \cite{Bloom_Effects_1957}, giving
\begin{equation}
    \frac{d}{dt}\mathbf{M} = \gamma \mathbf{M} \times \mu_0 \left(\mathbf{H}_0 + \mathbf{H}_r\right)
\end{equation}
which gives the equations of motion
\begin{subequations}
\begin{align}
    \frac{d}{dt}M_x &= \mu_0 \gamma M_y H_0 + 2\pi Q\eta M_x M_z \\
    \frac{d}{dt}M_y &= -\mu_0 \gamma M_x H_0 + 2\pi Q\eta M_y M_z \\
    \frac{d}{dt}M_z &= -2\pi Q \eta \left(M_x^2 + M_y^2 \right)
\end{align}
\end{subequations}

Radiation damping leads to nonlinear equations of motion which are quite different from the equations which govern the magnetization in the dressed nutation phase, in the presence of a demagnetizing field (Equation 2).

\section{FID Phase}


Tip-angle dependent frequency shifts in the dressed Rabi oscillation phase are accompanied by tip-angle dependent phase shifts. There are two effects which contribute to this phase shift: magnetization dynamics during the pulse, and the change in spin ensemble precession frequency after the pulse. The first effect can be understood qualitatively by considering that the instantaneous effective field felt by the spin ensemble changes during the pulse, so that the instantaneous direction about which the magnetization vector precesses also changes during the pulse. Numerical simulations of Equation 2, such as those presented in Figures 1e and 2e in the manuscript, show how the magnetization vector can pick up a phase shift relative to the drive pulse.  The phase shift is particularly striking in Figure 1e, where a 20 ns pulse, which begins rotating the magnetization \textit{about} the y-axis, ends up with the magnetization nearly \textit{aligned with} the y-axis for a phase shift of nearly $\pi/2$. The second effect arises due to the fact that the magnetization picks up a phase after the microwave pulse, but before and during the detection window, which is proportional to the tip-angle dependent precession frequency, leading to an additional frequency shift which is sensitive to the time delay between the end of the microwave pulse and the beginning of the detection window.

To first order, the phase $\phi(t)$ acquired by the spin system in the rotating frame of the microwave pulse of frequency $\omega$ at a time $t$ after the start of the pulse can be estimated by
\begin{equation}
    \phi(t) = \int_0^t \left[\omega(M_z(t')) - \omega\right]dt' = (\omega_L - \omega)t -\int_0^t\mu_0\gamma\theta_d M_z(t')dt'
    \label{eq:phase_shift_during_deadtime}
\end{equation}
Accurately measuring this phase shift in FEL-EPR experiments is experimentally challenging due to the experimental deadtime between the end of the high power FEL pulse and the beginning of data acquisition. This deadtime is limited by FEL ring-down and by scattered light which can partially saturate the detector. After the microwave pulse but before detection begins, the spin ensemble picks up a phase shift according to Equation \ref{eq:phase_shift_during_deadtime}, which is large for typical experimental deadtimes of $\sim70$ ns. The phase acquired during the experimental deadtime therefore dominates the phase of the measured FID.

\begin{figure}[ht]
    \centering
    \includegraphics{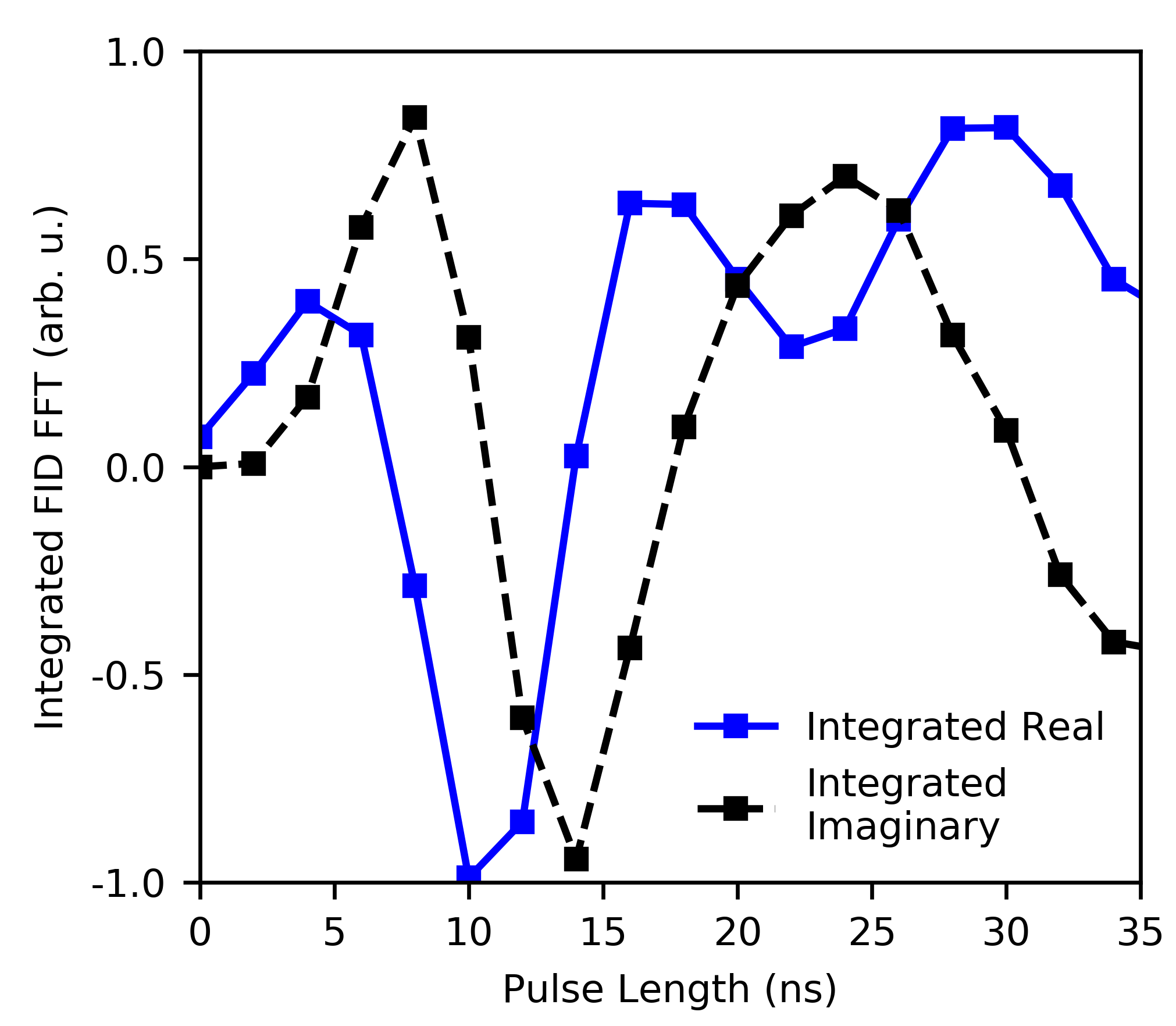}
    \caption{
    Integrated real and imaginary components of the FT-EPR lineshape as a function of pulse length, for the data shown in Figure 1c of the manuscript. 
    }
    \label{fig:IntegratedFTEPR_Figure1}
\end{figure}

\begin{figure}[ht]
    \centering
    \includegraphics{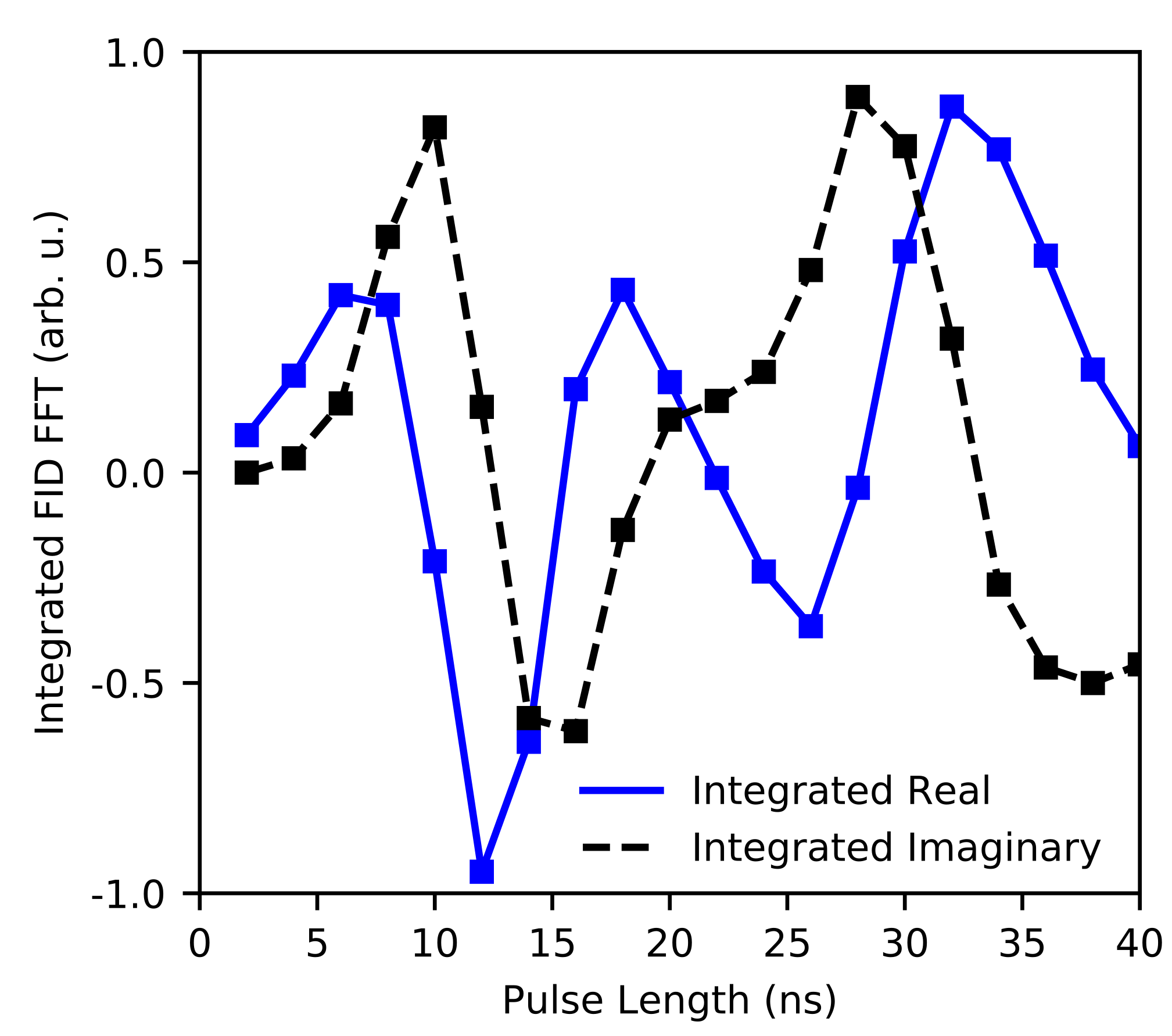}
    \caption{
    Integrated real and imaginary components of the FT-EPR lineshape as a function of pulse length, for the data shown in Figure 2 of the manuscript. 
    }
    \label{fig:IntegratedFTEPR_Figure2}
\end{figure}

Because we were unable to measure this phase shift, the integrated power spectrum, rather than the integrated absorption lineshape, was used in Figures 1 and 2 of the manuscript, in order to clearly show the change in FID amplitude with pulse length. The integrated complex FT-EPR lineshape for the data shown in Figures 1 and 2 are presented in Figures \ref{fig:IntegratedFTEPR_Figure1} and \ref{fig:IntegratedFTEPR_Figure2}, respectively.

When the FT-EPR absorption lineshape is shown, as in Figures 1b, 1d, 2b, and 4a-c, a phase shift has been applied to the FT-EPR lineshape so that the absorption lineshape is positive. 

\section{References}
\label{References}

\bibliography{biblio}